\documentclass[12pt]{article}
\usepackage{amsmath}
\usepackage{graphicx}
\usepackage{natbib} 
\usepackage{comment}
\newcommand{\blind}{0}

 \usepackage{amsfonts}
 \usepackage{times}
 \usepackage{bm}
 \usepackage{xcolor}
 \usepackage[plain,noend]{algorithm2e}

\newtheorem{corollary}{Corollary}

\def\v{{\varepsilon}}
\def\EVOIR{\textsc{evoir}}

\def\VSI{\textsc{vsi}}
\def\RVSI{\textsc{rvsi}}
\def\PVSI{\textsc{pvsi}}

\begin{document}

\def\spacingset#1{\renewcommand{\baselinestretch}%
{#1}\small\normalsize} \spacingset{1.3}
\if0\blind
{
  \title{\bf A Unified Approach for Outliers and Influential Data Detection  -- The Value of Information in Retrospect}
  \author{Jacob Parsons and Le Bao \thanks{This project was supported by NIH/NIAID R01AI136664.}\\
    Department of Statistics, Penn State University,\\ University Park, PA, USA \\
    lebao@psu.edu}
  \maketitle
} \fi

\if1\blind
{
  \bigskip
  \bigskip
  \bigskip
  \begin{center}
    {\LARGE\bf A Unified Approach for Outliers and Influential Data Detection  -- The Value of Information in Retrospect}
\end{center}
  \medskip
} \fi

\bigskip

\begin{abstract}
Identifying influential and outlying data is important as it would guide the effective collection of future data and the proper use of existing information. We develop a unified approach for outlier detection and influence analysis. 
Our proposed method is grounded in the intuitive value of information concepts and has a distinct advantage in interpretability and flexibility when compared to existing methods: it decomposes the data influence into the leverage effect (expected to be influential) and the outlying effect (surprisingly more influential than being expected); and it applies to all decision problems such as estimation, prediction, and hypothesis testing. 
We study the theoretical properties of three value of information quantities, establish the relationship between the proposed measures and classic measures in the linear regression setting, and provide real data analysis examples of how to apply the new value of information approach in the cases of linear regression, generalized linear mixed model, and hypothesis testing.
\end{abstract}

\noindent%
\vfill

\newpage
\spacingset{1.3} 

\section{Introduction}
\label{sec:intro}
In the course of any statistical analysis, it is necessary to consider issues of data quality and model appropriateness. To this end, it is helpful to identify influential and outlying data. 
A portion of the data is influential if its inclusion causes the fit of a model to shift substantially. When checking the quality of data, the focus should be placed on investigating data that have the largest impact on any decisions that are to be made based on the data. A portion of the data is outlying if it is very distant from what would be predicted from the model using the rest of the data. Outlying data are important in checking data quality as it may indicate that a portion of the data is more likely to have quality issues. The presence of outlying data may also suggest that the model being used is inappropriate. 

The diagnostic tools for influential data points and outliers have generated a very large body of work in the context of linear regression models \citep{Cook1982residuals,Belsley2005regression,Chatterjee2009sensitivity}. The leverage score, the externally studentized residuals, Cook’s distance, and the Welsch-Kuh distance (also called DFFITS) are just a few of many frequentist metrics. 
\cite{Chatterjee1992review} discuss the distinction between outliers and influential points. For instance, Cook’s distance or Welsch-Kuh distance identifies the influential data point, and the distance can be decomposed into the studentized residuals (the outlying effect due to unusually extreme values in the response variables) and the leverage (extreme values in predictors) \citep{Belsley2005regression}. However, the distinction between outliers and influential points is rarely discussed beyond the linear regression models. 

\citet{Johnson1983predictive,Johnson1985estimative,pettit1985,carlin1991,Guttman1993bayesian} develop influential case detection methods for linear models from a Bayesian viewpoint. Those early approaches measure the influence of a portion of the data using the Kullback--Leibler (KL) divergence between the posterior distributions calculated based on all of the data and the posterior distribution that results from excluding the portion of the data under consideration. Other forms of divergence measures are considered by \citet{Peng1995bayesian}. \citet{ali1990} presents an approach to influence analysis based on the measure of average information suggested by \citet{lindley1956}. 
 \citet{Mcculloch1989local} takes a local influence approach to model perturbation and quantifies influence by the curvature of the KL divergence (between perturbed and unperturbed posteriors). With a similar idea, \citet{Millar2007assessment} evaluate the local influence of conditional independent observations by using a geometrically weighted likelihood, and simplify the second derivative of the KL divergence to the posterior variance of the log-likelihood. \citet{Van2007local} generalize it to multiplicative modes of perturbation. Various perturbation schemes to the data are discussed by \citet{zhu2011}. 
Several approaches to identifying outliers in a Bayesian setting have also been proposed previously, most of which define the outlyingness based on the posterior predictive distribution, for example, \cite{Box1980sampling,Geisser1980discussion,pettit1985,Johnson1983predictive,Johnson1985estimative,Geisser1985prediction,Geisser1987influential,Guttman1988outliers,Pena1993comparing,Guttman1993bayesian}. The posterior predictive check becomes one of the most popular Bayesian model diagnosis tools \citep{Gelman1996posterior,Gelman2006data}. \cite{Chaloner1988bayesian} define an outlier as an observation with an extreme random error based on the posterior probabilities of regression error terms \citep{Zellner1975bayesian}. While all of these Bayesian methods use different influence measures or outlier detection metrics, none of these methods discusses the distinction or connection between influential case detection and outlier detection, except for \citet{weiss1996} who explains the need for both influence and outlier statistics in case-influence analysis.

The value of information (VOI) method was initially put forward in the middle of the twentieth century during the development of statistical decision theory. It was designed to help in deciding if an experiment is worth conducting, choosing between different research regimes, and determining optimal sample size \citep{raiffa1961}. However, since their genesis, VOI methods have been largely neglected by statisticians. They have, however, seen success in many applied settings. \citet{keisler2014} provide a good summary of the applied work that has been done using VOI methods. 

In this article, we propose a unified approach for identifying outliers and influential data for general models using VOI concepts. Our contributions to the literature are: (1) we extend VOI methods to applications outside of planning experiments such as outlier and influential data diagnostics; (2) the measure for the data influence is decomposed into the leverage effect (expected to be influential) and the outlying effect (surprisingly more influential than expected); (3) it applies to all decision problems; and (4) the influence measure is directly related to the goal of the analysis.


In the next section, we introduce concepts, notations, and quantities relating to the value of sample information. In Section~\ref{sec:voiinfluence}, we establish a connection between the expected value of sample information and a decision-theoretic influence measure first put forward by \citet{kempthorne1986} and propose the use of the expected value of information ratio. 
In Section~\ref{sec:examples}, we present the closed-form expressions of the value of information measures for linear regression and quadratic losses. We also demonstrate that the proposed approach is generic and applicable to settings outside conjugate priors and different loss functions with classification and hypothesis testing examples.
Section~\ref{sec:computation} shows how to estimate the value of information measures using Monte Carlo methods and approximation methods. 
In Section~\ref{sec:application}, we illustrate the use of the proposed quantities using two data sets: the U.S. employment regression model that was originally used in the development of Cook's distance; and the HIV prevalence estimation in the Kingdom of Eswatini; we also present an approximation
approach when the explicit forms for the value of information measures are unavailable. In Section~\ref{sec:discussion}, we offer conclusions and discussion for future work. 


\section{The Value of Sample Information}
\label{sec:voi}

In decision theory, the goal is to choose the best action $a$ from a set of possible actions $\mathcal{A}$ called the action space. Typically, how good an action is depends on some unobserved parameter $\theta$ taking values in a parameter space $\Theta$.  We quantify how good or bad an action is using a loss function, $L(a, \theta) : \mathcal{A} \times \Theta \to \mathcal{R}$, whose value depends on both the parameter $\theta$ and the action $a$. The larger the loss, the less preferable the action. 

We typically do not know the true value of $\theta$. So, we must choose the optimal action while taking into account the uncertainty about $\theta$. In the Bayesian setting, we may choose an action by minimizing the expected loss for an action conditional on all of the information that is available to us. The resulting choice is called the Bayes action. For instance, if we have not yet collected any data, then the Bayes action is the $a_0 \in \mathcal{A}$ that minimizes the prior risk $$R(a) = E \big \{ L(a,\theta) \big \} =  \int  L(a, \theta) d P (\theta).$$ After observing data $Y$, the Bayes action is the $a_Y$ that minimizes $$R(a \mid Y) = E \big \{ L(a,\theta) \mid Y \big \} = \int L(a, \theta) dP(\theta \mid Y)$$ where $P(\theta)$ is the prior distribution and $P(\theta \mid Y)$ is the posterior distribution. We shall use a similar notation when observing multiple observations (e.g. the Bayes action after observing two sets of observations $Y_1$ and $Y_2$ shall be denoted $a_{Y_1,Y_2}$). 

The Bayes action made conditional only on prior information, $a_0$, provides a baseline level of loss to compare with the action that makes use of additional information $Y$. In particular, the value of sample information ($\VSI$) that a sample $Y$ provides to an actor with some imperfect information summarized in a prior distribution is the reduction in loss that would occur if an action is made conditional $Y$ rather than just prior information, $$\VSI(Y, \theta) = L(a_0, \theta) - L(a_Y, \theta).$$ Having already observed data $Y_1$, the additional value of sample information provided by the additional data $Y_2$ for the actor is the additional reduction in loss that occurs if the decision is made conditional on both $Y_1$ and $Y_2$ rather than just $Y_1$, $$\VSI(Y_2 \mid Y_1; \theta) = L(a_{Y_1}, \theta) - L(a_{Y_1,Y_2}, \theta).$$ 
As the true value of $\theta$ is unknown, we have to take the expectation of $\VSI$ over possible values of $\theta$.
In the next section, we will use multiple expectations of the $\VSI$, each conditioned on a different set of information. 





\section{Evaluating Influence Using The Expected Value of Information}
\label{sec:voiinfluence}


Previous works in the value of information have primarily been concerned with the $\VSI$ that take expectations conditional on all currently available data \citep{keisler2014, heath2017}. The primary goal has been to evaluate potential new sources of information or establish bounds on the benefit of collecting certain kinds of data. A potential data source is only thought to be worth obtaining if the expected value of sample information corresponding to the source is greater than its cost. 

Our proposal goes beyond evaluating the contribution of future data. We also use $\VSI$ as a diagnostic tool to analyze the contribution of various data that have already been collected. \citet{kempthorne1986} suggests using the expected increase in the loss that would be incurred by incorrectly excluding a data point from a linear regression model as one measure of influence. We think the idea is generally applicable to any setting that can be formulated as a decision problem. 

Suppose the data can be partitioned into portions $Y_1, ..., Y_N$, and one is interested in the influence of a portion of the data $Y_i$, then the expectation of $\VSI$ is 
$$E \big \{ \VSI(Y_i \mid Y_{-i} ; \theta)\big \} = E \big \{ L(a_{Y_{-i}}, \theta)  - L(a_{Y_{-i}, Y_i}, \theta) \big \},$$
where $Y_{-i}$ is the data with the portion $Y_i$ removed, $E \{ L(a_{Y_{-i}}, \theta)\}$ is the risk of the action $a_{Y_{-i}}$, and $E \{ L(a_{Y_{-i}, Y_i}, \theta)\}$ is the risk of the action $a_{Y_{-i}, Y_i}$.

To differentiate between two versions of $E \big \{ \VSI(Y_i \mid Y_{-i} ; \theta)\big \}$ under two states of information, with and without knowing $Y_i$, we shall call $\PVSI(Y_i \mid Y_{-i} ; \theta) = E \big \{ \VSI(Y_i \mid Y_{-i} ; \theta) \mid Y_{-i} \big \}$ the prospective expected value of sample information ($\PVSI$) where $Y_i$ is the future data, and $\RVSI(Y_i \mid Y_{-i} ; \theta) = E \big \{\VSI(Y_i \mid Y_{-i} ; \theta) \mid Y_{-i}, Y_i  \big \}$ the retrospective expected value of sample information ($\RVSI$) where $Y_i$ has been observed. 

Using only $\RVSI$, we can identify the points that have had the most influence on the decision under consideration, but the scale of this measure is not always clear. In particular, we cannot say if the influence of a portion of the data is larger than expected or if it is due to the amount of information that the data source brings. However, $\PVSI$ tells us how much influence we should expect a portion of the data to have on the decision. The law of total expectation relates the retrospective $\VSI$ to the prospective $\VSI$ as follows: 
\begin{equation}
  \label{Equation::proRetRel}
  E \Big [ \RVSI(Y_i \mid Y_{-i} ; \theta) \mid Y_{-i} \Big ] = \PVSI(Y_i \mid Y_{-i} ; \theta).
\end{equation}

A natural comparison between the observed influence and the expected influence of a portion of data is the expected value of information ratio, $\EVOIR$,
$$
\EVOIR(Y_i \mid Y_{-i}) = \frac{\RVSI(Y_i \mid Y_{-i} ; \theta)}{\PVSI(Y_i \mid Y_{-i} ; \theta)} = \frac{E\{\VSI(Y_i \mid Y_{-i} ; \theta) \mid Y_{-i}, Y_i\}}{ E \big \{ \VSI(Y_i \mid Y_{-i} ; \theta) \mid Y_{-i} \big \}}.
$$
Following from Equation \ref{Equation::proRetRel}, 
A very coarse interpretation of this ratio is that
$
E \big \{ \EVOIR(Y_i \mid Y_{-i}) \mid Y_{-i} \big \} = 1.
$
Thus, a portion of the data with an expected value of information ratio greater than one is more influential than expected based on the rest of the data. 


It is useful to know the distribution of $\EVOIR$ as it allows one to measure exactly how often we should expect to see as large of an influence as we do. Since the influence of $Y_i$ is of interest, $Y_{-i}$ shall always be fixed, and the uncertainty of $\EVOIR$ comes from the predictive distribution of $Y_i|Y_{-i}$. The $p$-value of $\EVOIR$ is defined as the conditional probability of observing another $Y_i$ from the distribution of $Y_i|Y_{-i}$ that is equally or more influential than the observed $Y_i$. 

Next, we present the analytical distributions of $\EVOIR$ under the quadratic loss function and the linear regression model in Sections~\ref{sec:quadratic} and ~\ref{sec:regression}. When an analytical distribution of $\EVOIR$ is unavailable, we can approximate it numerically through Monte Carlo methods where the observed $Y_i$ is replaced by the Monte Carlo samples generated from the predictive distribution of $Y_i|Y_{-i}$. 
Monte Carlo approximation are presented in Section~\ref{sec:computation} and examples are shown in Sections~\ref{sec:LongleyTesting} and~\ref{sec:swaziland}.


\section{Examples of Value of Information Measures}
\label{sec:examples}
In Section~\ref{sec:quadratic}, we present the closed-form expressions of $\RVSI$, $\PVSI$, and the distribution of $\EVOIR$ under the quadratic loss functions. An analogy exists between these measures and quantities used in frequentist influence analysis for linear regression, as we shall see in Section~\ref{sec:regression}. To illustrate the wider applicability of the proposed value of information framework for influence analysis, we apply these methods to the problems of classification and hypothesis testing in Section~\ref{sec:testing}

\subsection{Value of Information Under the Quadratic Loss Function}
\label{sec:quadratic}

Consider an estimation problem in which $\Theta = \mathcal{A} = \mathbb{R}^p$. A quadratic loss function has the advantages of being computationally efficient, interpretable, and familiar:
$$
L(a, \theta) = (a - \theta)^TQ(a-\theta)
$$
where $Q = A^TA \in \mathbb{R}^{p \times p}$ is positive definite. In this situation, the loss function is just the squared distance between an estimate and the true value of the parameter using the metric defined by $Q$, which is known as the Mahalanobis distance \citep{Mahalanobis1936}. It is often easier to interpret $A$ than it is to interpret $Q$. $A$ can be thought of as a linear transformation of the parameter space into a space that is more appropriate to measure distances. For instance, when measuring prediction errors in linear regression, a design matrix $X$ can be used to transform the coefficient vector $\beta$ into a vector of predicted values and $Q = X^TX$. The Bayes action given data $Y$ is $a_Y = E(\theta \mid Y)$, the posterior mean of $\theta$. The law of total expectation yields  
$$
a_{Y_{-i}} = E \big \{ E(\theta \mid Y_{-i} , Y_i) \mid Y_{-i} \big\} = E(a_{Y_{-i}, Y_i} \mid Y_{-i} ).
$$ 
In this situation, the retrospective value of sample information is given in the following corollaries:

\begin{corollary}
\label{thm:retrospectiveForm}
Let $\theta$ be a $p$-dimensional parameter. Suppose that $Y_{-i}$, $Y_i$ are random objects defined on the same sample space with distributions depending on $\theta$. Then, if $\mathcal{A} = \mathbb{R}^p$, $L(a, \theta) = (a - \theta)^TQ(a-\theta)$ where $Q = A^TA  \in \mathbb{R}^{p \times p}$ is positive definite, and the distributions of $\theta \mid Y_{-i}$ and $\theta \mid Y_{-i}, Y_i$ are proper with finite means:
\begin{equation*}
    E \big \{\VSI(Y_i \mid Y_{-i}; \theta)  \mid Y_{-i}, Y_i \big \} = (a_{Y_{-i}} - a_{Y_{-i},Y_i})^T Q (a_{Y_{-i}} - a_{Y_{-i},Y_i}).
\end{equation*}
\end{corollary}

Corollary~\ref{thm:retrospectiveForm} tells us that the retrospective value of information that is used to measure the influence of $Y_i$ is how far the estimate moved in the transformed parameter space by including $Y_i$ in the analysis in addition to $Y_{-i}$. We also have the following result:
\begin{corollary}
  \label{thm:prospectiveForm}
  Under the assumptions of Corollary~\ref{thm:retrospectiveForm}, the prospective expected value of sample information is 
  $$
  E \big \{ \VSI(Y_i \mid Y_{-i}; \theta)  \mid Y_{-i} \big \} = tr\Big\{var \big ( A a_{Y_{-i},Y_i}  \mid Y_{-i} \big) \Big\}.
  $$
\end{corollary}

So, the expected influence of $Y_i$ given the rest of the data, or equivalently the expected squared distance between the two estimates made with and without $Y_i$, is the sum of the conditional variances of each component of the Bayes estimator after applying the transformation corresponding to $A$. We can sometimes give a finer-grained interpretation of the expected value of information ratio using the following fact:

\begin{corollary}
  \label{thm:evoir}
  Under the assumptions of Corollary \ref{thm:retrospectiveForm} and the additional assumption that $a_{Y_{-i},Y_i} \mid Y_{-i} \sim N(a_{Y_{-i}}, c \Sigma)$ such that $c > 0$ and $\Sigma Q$ is idempotent. Then, 
  $$
  \EVOIR(Y_i \mid Y_{-i}) \sim \chi^2_p / p, 
  $$	
  where $p = trace(\Sigma Q) = rank(\Sigma)$.
\end{corollary}

This last corollary could be useful in establishing the closed-form p-values for interpreting how extreme a value of the $\EVOIR$ is under certain circumstances, e.g., normal linear regression models. In particular, this corollary applies to the regression examples presented in the next section when the variance is known, and the $\EVOIR$ follows an F-distribution when the variance is unknown. See Supplement S1 for the derivations of corollaries 1$\sim$3.

\subsection{Value of Information for Linear Regression}
\label{sec:regression}
In this section, we develop the value of information approach to influence analysis in the linear regression setting in order to illustrate how to use and interpret the proposed measures. Let $Y$ be an $n$-dimensional random vector such that 
\begin{equation}
Y \mid \beta, \sigma^2 \sim N\big( X \beta , \sigma^2 I_n \big),
\label{eqn:lm}
\end{equation}
where $X$ is a $n \times p$ matrix, and $\beta$ is a $p$-dimensional vector, and $\sigma^2 \in \mathcal{R}$. We assume that we observe $Y$ and that $X$ is known, but that $\beta$ and $\sigma^2$ cannot be observed directly.

\textbf{Non-informative Prior:}
We first assign a non-informative prior distribution to $\beta$ and $\sigma^2$:  
$\pi (\beta, \sigma^2) \propto \sigma^{-2}$. Then, 
$$
\beta \mid \sigma^2, Y \sim N\Big\{\hat{\beta} , \sigma^2 \big( X^T X \big)^{-1} \Big\}, \quad \sigma^2 \mid Y \sim \chi^{-2} (n-p, S^2),
$$ 
where $\hat{\beta} = \big( X^T X \big)^{-1} X^T Y$
is the maximum likelihood estimate for $\beta$ and $S^2 = \frac{1}{n-p} (Y - X \hat{\beta} )^T(Y  - X \hat{\beta})$
is the standard unbiased estimate for the error variance. See for instance \citet{bda}. We will also make use of the symmetric hat matrix $H = X (X^TX)^{-1} X^T$ 
with entries $h_{ij}$. Its diagonal entries $h_{ii}= X_i (X^TX)^{-1} X_i^T$ are known as the leverage of the $i$th observation, where $X_i$ is the $i$th row of the matrix $X$.

We are interested in measuring the influence of each observation on predictions for the mean value of a new observation $Y_{new} \sim N(X_{new} \beta, \sigma^2)$. In this situation we would like to choose an action $a \in \mathcal{R}^p$ that minimizes $E(X_{new}a - X_{new}\beta)^2$. Unfortunately, this would require us to either specify a particular $X_{new}$ or specify a distribution of $X_{new}$. Choosing a particular $X_{new}$ would be overly restrictive, and in general, we would rather not specify a particular form for the distribution of the independent variables in a linear regression setting. With this in mind, we will assume that $X_{new}$ comes from the empirical distribution of the rows of $X$. This suggests using the following loss function:
$$
L(a, \beta) = (Xa - X\beta )^T (Xa - X\beta ) = (a - \beta)^T X^TX (a - \beta).
$$
The Bayes action based on $Y$ is $\hat{\beta} = E(\beta \mid Y)$. To evaluate the influence of the $i$th observation, we introduce the following corollaries.

\begin{corollary}
\label{thm:lm1}
Assume the normal linear regression model in Equation~\ref{eqn:lm} and $X^TX$ is positive definite. Let $Y_{-i}$ be the reduced data obtained by removing $Y_i$ from $Y$. Let $X_{-i}$ denote the matrix obtained by removing the $i$th row $X_i$ from the matrix of predictors $X$. The retrospective expected value of sample information ($\RVSI$) is 
$$
  E \big \{ \VSI(Y_i \mid Y_{-i}; \beta)  \mid Y \big \} = \sum_{k = 1}^n (X_k \hat{\beta} - X_k \hat{\beta}_{-i})^2,
$$
where $\hat{\beta}_{-i}$ is the maximum likelihood estimate for $\beta$ based on the reduced data $Y_{-i}$ and $X_{-i}$.
\end{corollary}

The above $\RVSI$ of $Y_i$ is an unscaled version of the Cooks distance for the $i$th data point: $\sum_{k = 1}^n (X_k\hat{\beta} - X_k\hat{\beta}_{-i})^2/pS^2.$
Cook's distance is a common frequentist measure of influence in linear regression \citep{cook1977}. Notice that the scaling factor is the same for all data points in the sample. So, we will draw the same conclusions about the relative influence or value of points using either the Cook's distance or $\RVSI$. 

\begin{corollary}
\label{thm:lm2}
Under the assumptions of Corollary~\ref{thm:lm1}, the prospective expected value of sample information ($\PVSI$) is 
$$
E \big \{ \VSI(Y_i \mid Y_{-i}; \beta)  \mid Y_{-i} \big \} = \frac{n-p-1}{n-p-3} S_{-i}^2 \frac{h_{ii}}{1 - h_{ii}},
$$
where $S_{-i}^2$ is the unbiased estimate for the variance based on $Y_{-i}$ and $X_{-i}$.
\end{corollary}

This $\PVSI$ plays a role similar to the leverage in the frequentist setting in that both can be used to measure how influential we would expect an observation to be according to the model without observing the actual response. We see that $\PVSI$ is an increasing function of the leverage but also depends on the sample variance $S^2_{-i}$. If the sample size is sufficiently large, $S^2_{-i}$
will be similar for all $i$ and sorting the points according to their prospective value of sample information would give the same ordering of points as if we had sorted them by their leverage.

\begin{corollary}
\label{thm:lm3}
Under the assumptions of Corollary~\ref{thm:lm1}, the expected value of information ratio is $\EVOIR(Y_i \mid Y_{-i}) =  \frac{(n-p-3) }{(n-p-1)}t_{(i)}^2,$
where $t_{(i)}$ is the externally Studentized residual for the $i$th observation, and the distribution of $t_{(i)}^2$ is an F-distribution having $1$ and $n-p-1$ degrees of freedom.
\end{corollary}

The $\EVOIR$ is therefore large when it is far from the line predicted by the other points as measured by the externally Studentized residual. See Supplement S2 for the derivations of these corollaries.

\textbf{Informative Prior:}
To understand the effect of prior information in the value of information approach, we also consider the informative $g$-prior for $\beta$ \citep{Zellner1986assessing} and the conjugate prior for $\sigma^2$:
$$
\begin{aligned}
\beta &\sim \textup{normal}(\beta_0, g \sigma^2 (X^TX)^{-1}),\\
\sigma^2 &\sim \textup{inverse-gamma}(\nu_0/2, \nu_0\sigma_0^2/2).
\end{aligned}
$$ 
The $g$-prior is widely used for regression parameters because the resulting posterior distribution of $\beta$ is invariant to changes in the scale of the covariates. For example, in the application of Longley data analysis that we present in Section \ref{sec-application-lm}, time is an important regressor. It makes sense that the posterior expected change in the response for a year change in time is the same, whether the time is recorded in terms of months or years. Derivations of $\RVSI$, $\PVSI$, and $\EVOIR$ under the $g$-prior are provided in Supplement S3.

\subsection{Value of Information for Hypothesis Testing}
\label{sec:testing}

One could approach the problems of classification and hypothesis testing from a Bayesian decision theory framework. The set of possible classifications or hypotheses are the action space. Let $c_1,...,c_K$ be the set of possible class assignments to a set of objects (or labels for potential hypotheses); and $\theta$ be the correct classification of the objects under consideration or the true hypothesis. For evaluating the influence of data, it is preferable to consider not just the decision that is made but also the degree of certainty with which it is made.
So, instead of selecting a single assignment of classes, we assign a probability to each potential assignment of classes. Therefore, we take the action space to be the $K$-dimensional probability simplex:
$\big \{p \in [0,1]^K : \sum \limits_{k =1}^{K} p_k = 1 \big \}.$

A common loss function for this continuous treatment of classification is the cross entropy loss: $L(p, \theta) = - \sum \limits_{k=1}^{K} I\{\theta = c_k \} \log (p_k),$
in which the probability distribution expressed by $p$ is used to summarize the current level of knowledge. The prior expected loss also takes the form of a cross entropy:
$E[L(p, \theta)] = - \sum \limits_{k =1}^{K} P(\theta = c_k) \log (p_k).$
The Bayes action under only prior minimizes the above risk by matching the prior probability distribution over the values of $\theta$: $p_0 = \big [ P(\theta = c_1), \dots, P(\theta = c_K) \big].$
In a like manner, the Bayes action given data $Y$ is: $p_Y = \big [ P(\theta = c_1 \vert Y), \dots, P(\theta = c_K \vert Y) \big].$ It follows that $\RVSI$ is the Kullback–Leibler (KL) divergence between the posterior distribution conditional on $Y_{-i}$ and the posterior distribution conditional on both $Y_{-i}$ and $Y_i$:
$$
E\{\VSI(Y_i \mid Y_{-i} ; \theta) \mid Y_{-i}, Y_i \} = -\sum \limits_{k=1}^K P\big(\theta = c_k \vert Y_{-i}, Y_i\big) \log \frac{P\big(\theta = c_k \vert Y_{-i} \big)}{ P\big(\theta = c_k \vert Y_{-i}, Y_i\big)}.
$$

Then, $\PVSI$ 
integrates $\RVSI$ over $Y_i \vert Y_{-i} $:
$$
\begin{aligned}
E\{\VSI(Y_i \mid Y_{-i} ; \theta) \mid Y_{-i} \} &= E \Big [ -\sum \limits_{k=1}^K P\big(\theta = c_k \vert Y_{-i}, Y_i\big) \log \frac{P\big(\theta = c_k  \vert Y_{-i} \big )}{ P\big(\theta = c_k \vert Y_{-i}, Y_i\big)} \Big \vert Y_{-i} \Big ].
\end{aligned}
$$
The above expected KL divergence can sometimes be expressed analytically but not generally so. In Section~\ref{sec:LongleyTesting}, we illustrate the numerical integration drawing samples of $Y_i$'s from its predictive distribution conditioning on $Y_{-i}$. 

\section{Approximation of Value of Information Measures} 
\label{sec:computation}
When the explicit forms of VOI measures are not available, we need to approximate them for each observation. Determining the retrospective expected value of sample information ($\RVSI$) is relatively straightforward and can be done using standard methods. 
We first obtain posterior samples conditional on all of the data and then obtain posterior samples conditional on the data with each observation removed in turn. Take the squared loss function for estimating the unknown parameter, $\theta$, as an example. The $\RVSI$ is computed for each observation by taking the squared distance between the posterior means based on the posterior conditional on the complete data and the data without that observation included. 


Computing the prospective expected value of sample information ($\PVSI$) is more computationally intensive. A naive Monte Carlo procedure for estimating the prospective expected value of sample information for the $i$th observation, $Y_i$, is given by the following:
\begin{enumerate}
\item Outer Monte Carlo: Sample $(Y_i^{(1)}, \theta^{(1)}), \dots, (Y_i^{(N)}, \theta^{(N)})$ from the joint conditional distribution of $Y_i, \theta \mid Y_{-i}$ giving a sample from the predictive distribution of the complete data and the corresponding true parameters.
\item Approximate the expected value,  $\theta_{ Y_{-i}} = E \big \{\theta \mid Y_{-i} \big \}$,  by $\hat{\theta} =  \frac{1}{N} \sum  \limits^{N}_{k=1} \theta^{(k)}$.
\item Inner Monte Carlo: Draw posterior samples of $\theta$ conditional on each set of complete data generated in step 1. That is, for $k = 1, \dots, N$, sample  $\theta^{(k,1)}, \dots, \theta^{(k,M)}$ from the conditional distribution of $\theta \mid Y_{-i}, Y_{-i}^{(k)}$.
\item Approximate the expected value conditional on each generated complete data sample $\theta_{ Y_{-i}, Y_i^{(k)}} = E \big \{ \theta \mid Y_{-i}, Y_i^{(k)} \big \}$ by  $\hat{\theta}^{(k)} =   \frac{1}{M} \sum  \limits^{M}_{j=1} \theta^{(k,j)}$.
\item Compute the $\PVSI$ based on the Monte Carlo sample: $E \big \{ \VSI(Y_i \mid Y_{-i}; \theta)  \mid Y_{-i}, Y_i \big \} \approx \frac{1}{N}  \sum \limits_{k=1}^N  (\hat{\theta} - \hat{\theta}^{(k)})^2$.
\end{enumerate}
Since sampling from a posterior distribution is often computationally intensive, and the number of samples required for a reasonable approximation grows very fast as the dimension of the parameter increases, this approach typically will not be feasible. To avoid the inner level of sampling in Step 3, we instead adopt the approaches of \citet{strong2015,Jackson2019value} and revise Step 3 and 4 of the previous procedure as follows:
\begin{enumerate}
\setcounter{enumi}{2}
\item Apply a non-parametric regression procedure to the pairs generated in (1) in order to estimate $f\{ Y_i^{(k)}\} = E  \big \{ \theta \mid  Y_{-i}, Y_i^{(k)} \big \}$.
\item Approximate $\theta_{ Y_{-i}, Y_i^{(k)}}$ by using the fitted values $\hat{\theta}^{(k)} = \hat{f} \{ Y_i^{(k)}\}$.
\end{enumerate}
We considered both a linear and a generalized additive model, as used by \citet{strong2015}, to estimate the functional form of $E\{\theta \mid Y^{(k)}\}$ in step 3, but further inspection showed substantial deviations between the approximation and the inner-level sampling method when the dimension of $\theta$ is high as in the HIV prevalence estimation example in Section~\ref{sec:swaziland}, where $\theta$ consists of HIV prevalence rates in multiple years and locations. In the end, we used a $K$-nearest-neighbor (KNN) regression to approximate $E\{\theta \mid Y^{(k)}\}$. In the KNN approach, the expected value of $\theta$ conditional on a simulated complete data set, $E \big \{ \theta \mid Y_{-i}, Y_i^{(k)} \big \}$, is taken to be the average of the $\theta^{(k')}$ whose corresponding $Y_i^{(k')}$ is near $Y_i^{(k)}$. Supplement 4 indicates a strong agreement between this approximation approach and the results that would occur by sampling from the posterior distribution conditional on the complete data directly to estimate the conditional means with high precision. Once $\RVSI$ and $\PVSI$ are computed, calculating the expected value of information ratio ($\EVOIR$) is trivially the ratio of the two.

\section{Applications}
\label{sec:application}
We illustrate our proposed approach using two data sets. The first data set contains employment rates, and other economic factors are used to provide an example of applying the new value of information approach in the case of linear regression, classification, and hypothesis testing. The second data set provides information about HIV prevalence in the Kingdom of Eswatini, formerly known as Swaziland. It is used to illustrate the use of a value of information approach to influence analysis in the case of a generalized linear mixed model. 

\subsection{Example: Linear Regression in Longley Data}
\label{sec-application-lm}
Initially, Cook (1977) illustrated the use of Cook's distance by an application to a data set first presented in by \citet{longley}. This data set contains the number of people employed in the United States and six other economic variables recorded from 1947 to 1962. In his example, Cook fit an ordinary linear regression having only first-order terms using the number of people employed as the response variable and all others as predictors. We shall take the same approach to illustrate how to interpret the measures discussed above. 

\begin{figure}[!ht]
\centering
\includegraphics[scale=.55,angle =270]{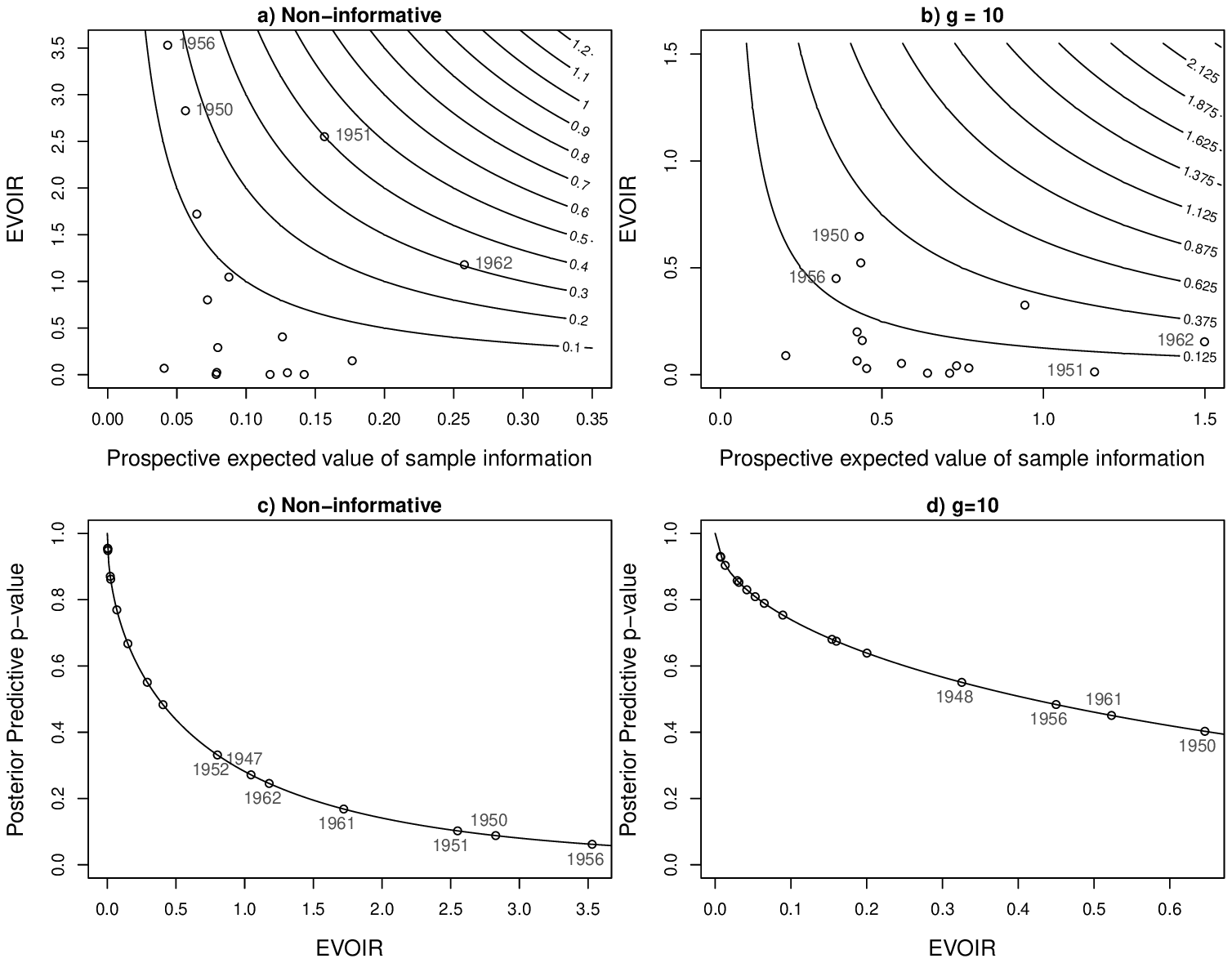}
\caption{The sub-figures in the first row present the expected value of information ratio plotted against the prospective expected value of sample information with contours indicating the retrospective expected value of sample information. 
The sub-figures in the second row plot the p-values of $\EVOIR$ under the theoretical predictive distribution conditional on the reduced data set plotted against the $\EVOIR$ for estimating the expected number of people employed in the United States under the non-informative prior.
The sub-figures (a) and (c) show these metrics computed under the non-informative prior for estimating the expected number of people employed in the United States, while sub-figures (b) and (d) indicate the same metrics with a g-prior. 
}\label{fig:retrospectivevalue-g-prior}
\end{figure}

Figure \ref{fig:retrospectivevalue-g-prior} (a) and (b) present the expected value of information ratio ($\EVOIR$) plotted against the prospective expected value of sample information ($\PVSI$) with contours indicating the retrospective expected value of sample information ($\RVSI$). We refer to this kind of plot as the VOI plot. Points that lie vertically higher indicate an observation that has influenced the model to a larger degree than expected based on the rest of the data. Points that lie farther to the right correspond to observations expected to have a larger influence. Finally, points that are closer to the top right of the plot correspond to more influential observations. Sub-figure (a) shows these metrics computed under the non-informative prior for estimating the expected number of people employed in the United States, while sub-figure (b) indicates the same metrics with a g-prior. 

We first learn a few things from Figure \ref{fig:retrospectivevalue-g-prior} (a) under the non-informative prior. The observation made in 1951 has the largest influence on the model's fit as measured by the $\RVSI$ (contours). It was the second year of the Korean conflict. The observation is more than two and half times as influential as expected based on the rest of the data. 
The observation made in 1962 is the second most influential point. After the market experienced decades of growth since the Wall Street Crash of 1929, the stock market peaked during the end of 1961 and plummeted during the first half of 1962, also known as the Flash Crash of 1962. The plot indicates, however, that this influence is close to what would be expected according to the rest of the model and the covariates. The extreme values of covariates made it a high leverage point as indicated by the $\PVSI$ (x-axis). Thus, it is influential but not surprisingly so. The observations made in 1950 and 1956 have a much larger impact than would have been expected. Despite the high $\EVOIR$, the observations made in 1950 and 1956 are still less influential than those made in 1951 and 1962. This is due to the two points having an especially low prospective value, a consequence of being low leverage points. The numeric values of Cook’s distance, $\RVSI$, $\PVSI$, and $\EVOIR$ are presented in Supplement 5.

Under the g-prior, the $\RVSI$ in Supplement Equation 1 decreases as the prior information becomes stronger ($g$ decreases). The $PVSI$ in Supplement Equation 2 goes to zero as $g\to 0$, but this does not occur monotonically.
Comparing Figure \ref{fig:retrospectivevalue-g-prior} (a) and (b), we see that $\PVSI$ actually increases when $g$ decreases from $\infty$ to $10$. It is because the estimate for the residual variance, $\hat{\sigma}_{-i,g}^2$, is increasing at a faster rate than the remaining terms are decreasing as the prior becomes stronger for some intermediate values of $g$. 
The initial large increase in $\PVSI$ and the decrease in $\RVSI$ substantially reduce the magnitude of the $\EVOIR$ for the majority of observations. Once the increase in the variance estimate slows down, the $\EVOIR$ begins to increase again, stabilizing at the limiting value given in Supplement Equation 3.

Under both the setting of the non-informative prior and the g-prior with unknown variance 
the distribution of the $\EVOIR$ for an observation is an F-distribution having $1$ and $d$ degrees of freedom conditional on the remaining observations (see Supplement 3 for a demonstration of this fact). Here $d = n-p-1$ in the non-informative setting and $d = n$ in the g-prior setting. Figure \ref{fig:retrospectivevalue-g-prior} (c) and (d) 
plot the $p$-values in the theoretical distribution against the $\EVOIR$s of the corresponding observations for the non-informative prior setting and the g-prior setting with $g=10$. This illustrates that the outlying observations can be evaluated not only crudely using $\EVOIR$ but also in terms of $p$-values.


\subsection{Example: Hypothesis Testing in Longley Data}
\label{sec:LongleyTesting}
To illustrate the use of the value of information methods for influence analysis in a hypothesis testing scenario, we consider testing if the coefficient corresponding to the year in the Longley data is negative or positive, that is, whether employment has a positive or negative time trend that is not explained by the other variables. Using the notation from Section \ref{sec:testing}, the probability that the year coefficient, taken as the final coefficient, $\beta_{7}$ is less than zero can be determined by the fact that conditional on $Y_{-i}$:
$$
\frac{\beta_{7} - \hat{\beta}_{7, -i}}{ \sqrt{(X^T X)^{-1}_{7,7} S^2_{-i}} } \Big \vert Y_{-i} \sim t_{n - p - 1} 
$$
The $\RVSI$ for this hypothesis test using a cross entropy loss function is given by
$$
\begin{aligned}
  E\{\VSI(Y_i \mid Y_{-i} ; \theta) \mid Y_{-i}, Y_i \} 
  &= - P\big( \beta_{7} < 0 \vert Y_{-i}, Y_i\big) \log \frac{P\big( \beta_{7} < 0 \vert Y_{-i} \big)}{ P\big( \beta_{7} < 0 \vert Y_{-i}, Y_i\big)} \\
  & \quad \quad \quad  - P\big( \beta_{7} \geq 0 \vert Y_{-i}, Y_i\big) \log \frac{P\big( \beta_{7} \geq 0 \vert Y_{-i} \big)}{ P\big(\beta_{7} \geq 0 \vert Y_{-i}, Y_i\big)}
\end{aligned}
$$
The $\PVSI$ can be calculated easily by use of a Monte Carlo sample of $Y_i$'s from the predictive distribution of $Y_i$ conditional on $Y_{-i}$ and then taking an average of the corresponding $\RVSI$ calculated using the above formula. This is straightforward as $Y_i \vert Y_{-i}$ can be sampled in three steps:
\begin{enumerate}
    \item Sample $\sigma^2 \sim \chi^{-2}(n-p, S_{-i}^2)$ 
    \item Sample $\beta \sim N \big(\hat{\beta}_{-i}, \sigma^2 (X_{-i}^T X_{-i} )^{-1} \big) $
    \item Sample $Y_i \sim N(X_i \beta , \sigma^2)$
\end{enumerate}
The above procedure can be reduced to a single step that samples $T_i$ from a t-distribution with $n-p-1$ degrees of freedom and sets $Y_i = X_i \hat{\beta}_{-i} + S_{-i} \sqrt{(1-h_{ii})^{-1}} T_i  $.

\begin{figure}[!ht]
\begin{minipage}[b]{0.45\textwidth}
\centering
{(a) VOI plot}
\includegraphics[width=\linewidth,angle =270]{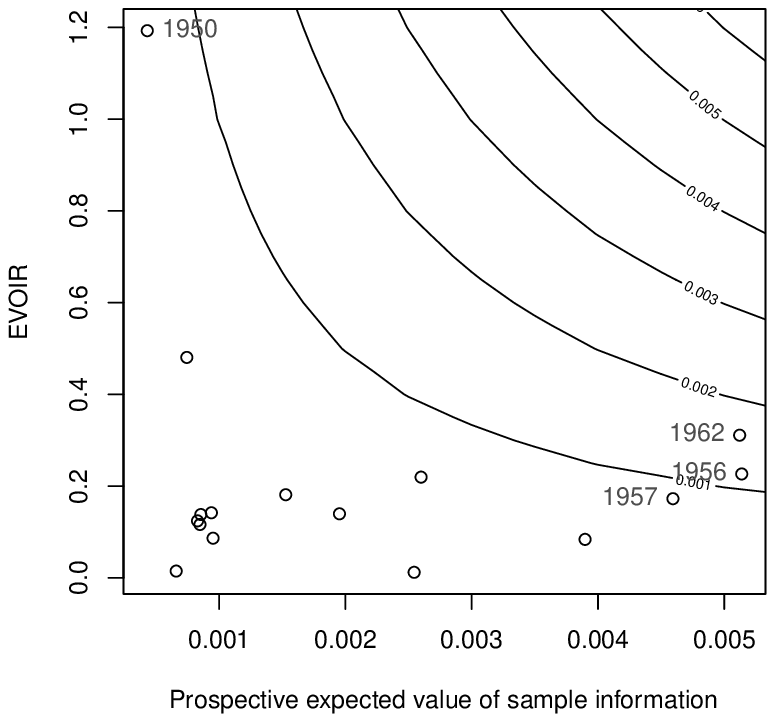}
\end{minipage}
\hspace{0.5cm}
\begin{minipage}[b]{0.45\textwidth}
\centering
{(b) $\PVSI$ plot}
\includegraphics[width=\linewidth,angle =270]{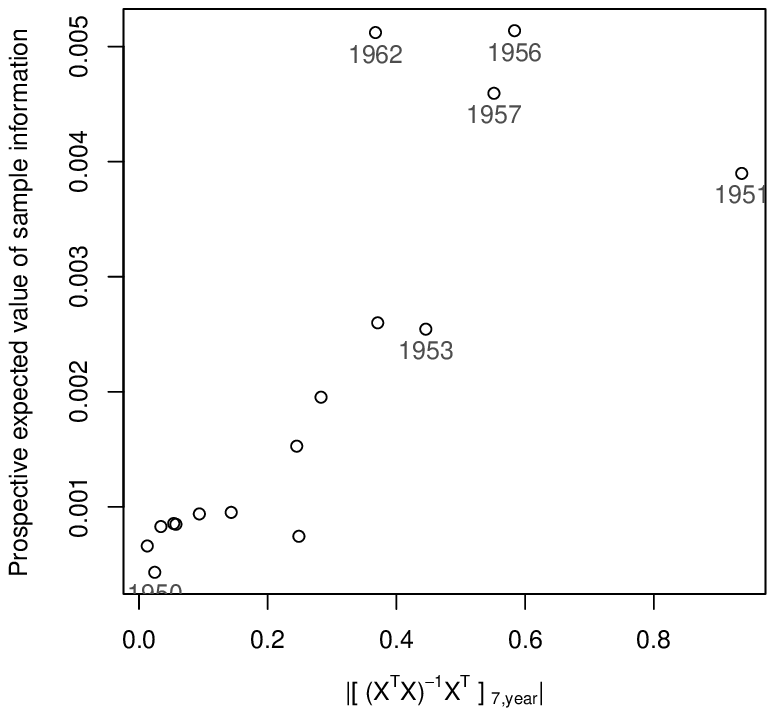}
\end{minipage}
\caption{The left sub-figure (a) shows the value of information plot for testing if the year coefficient is less than 0 for the Longley example under the non-informative prior. It shows the expected value of information ratio plotted against the prospective expected value of sample information, contours indicating the retrospective expected value of sample information. The right sub-figure (b) plots the prospective expected values sample information against $|[(X^TX)^{-1}X^T]_{7,year}|$.
}\label{fig:retrospectivevalue-testing}
\end{figure}

Figure \ref{fig:retrospectivevalue-testing} shows what results by applying this procedure to the Longley data using the same model as before. Only 1950 had an $\EVOIR$ greater than one, indicating that this is the only observation that had a larger than expected influence on choosing the sign of the year coefficient. A large number of points having the expected value of information ratios falling substantially below one may indicate problems with the model (recall that the mean of this quantity should be one). For instance, we may not be making use of all the information contained in the data or maybe neglecting correlation in the error terms.
The other data points with relatively large retrospective expected values of sample information ($\RVSI$) are from the years 1962, 1956, 1957, and 1951, mainly because of their large prospective expected values sample information ($\PVSI$). We further explore which factors affect $\PVSI$ the most. The sub-figure (b) shows that $\PVSI$ strongly depends on the absolute value of $[(X^TX)^{-1}X^T]_{7,year}$, where $(X^TX)^{-1}X^T$ is a $7 \times 16$ matrix with each entry corresponding to a covariate and year (7 covariates and 16 years in total). The sub-figure (b) plots $\PVSI$ against each entry of the 7th row of $(X^TX)^{-1}X^T$. Note that $X_7$ itself does not show as clear of a relationship with $(X^TX)^{-1}X^T$ (no single predictor does). These quantities reflect how the response is weighted for each year when estimating $\beta_7$ via maximum likelihood.

\subsection{HIV Prevalence in the Kingdom of Eswatini}
\label{sec:swaziland}
For more complicated models, the explicit forms for the value of information measures are often unknown. However, it is still possible to apply these methods numerically. In this section, we illustrate the use of these measures in such a setting by considering a generalized linear mixed model for HIV prevalence in the Kingdom of Eswatini.

\textbf{Background and Model:}
The Kingdom of Eswatini is a small developing country in Africa with a high occurrence of HIV. The main data source to inform estimates of HIV epidemics has been unlinked anonymous testing of pregnant women who attend antenatal clinics. Nearly all countries established antenatal clinics HIV surveillance in the early 1990s, making it the earliest and most consistently available source of information. The Kingdom of Eswatini has relatively sparse antenatal clinics data. Thus it is important to ensure that those data properly contribute to the estimation of the Kingdom of Eswatini HIV epidemic by detecting and investigating influential and outlying data.

For the purpose of estimating the prevalence of HIV in the country, patients at 17 different clinics were tested for the presence of HIV from 2002 to 2010, with data reported every two years. The Kingdom of Eswatini is comprised of four districts: Hhohho, Lubombo, Manzini, and Shiselweni. Five of the sites being monitored were in the Lubombo region, while each of the three remaining districts contained only four of the monitored clinics. One clinic in the Lubombo region reported no data for a single year, but otherwise, data exists for each clinic and period. We did not use the historical data before 2002 because they were not available at the local level, and we did not use any epidemiology model in this analysis. Therefore, the result only illustrates the value of information approach and should not be viewed as official HIV estimates for the Kingdom of Eswatini. 

Let $Y_{rst}$ be the number of individuals that test positive for HIV during the year $t$ at the $s$th site in the $r$th region. We assume that

$$
Y_{rst} \sim Binomial(N_{rst}, \pi_{rst}),
$$

where $\pi_{rst}$ is the HIV prevalence at the $s$th site in the $r$th region in year $t$, and $N_{rst}$ is the number of individuals tested for HIV at the $s$th site in the $r$th region in year $t$. Furthermore, we also assume that

$$
\pi_{rst} = \frac{1}{1 + \exp(-\eta_{rst})}, \quad \eta_{rst} = \mu + \alpha_r + f(t) + \gamma_s .
$$

The site effect is treated as a random effect with $\gamma_s \sim N(0, \tau^2)$, while the region effects $\alpha_r$ are fixed effects. The trend function $f(t)$ is approximated by a linear combination of cubic B-splines, giving rise to a vector $X(t) \in \mathbb{R}^3$ for each time point. That is, for some $\beta \in \mathbb{R}^3$,

$$
\eta_{rst} = \mu + \alpha_r + X(t)^T \beta + \gamma_s .
$$

We use assign weak independent prior distributions to the parameters:

$$
\begin{aligned}
  \mu &\sim N(0, 100)  \\
  \beta_i &\sim N(0, 100 ),\quad i = 1,\dots,3 \\
  \alpha_r &\sim N(0,100),\quad r = 2,3,4 \\
  \gamma_s &\sim N(0,\tau^2),\quad s = 1,\dots,17 \\
  \tau^2 &\sim Gamma^{-1}(0.1, 0.1).
\end{aligned}
$$

We set $\alpha_1 = 0$ for identifiability purposes, and each of the parameters is independent of the others in the prior distribution.

The main goal of the analysis is to estimate the prevalence of HIV for each of the four regions for each of the years examined. It is true that even according to the above model, each region has varying levels of HIV prevalence around each site. Thus we set the goal to be to estimate the following quantity for each region and year:

$$
\pi_{rt} = \frac{1}{1 + \exp(-\eta_{rt})}, \quad \eta_{rt} = \mu + \alpha_r + X(t)^T \beta.
$$

That is, we wish to estimate the matrix $\pi \in \mathbb{R}^{4 \times 5}$ whose entry in the $r$th row and $t$th column is $\pi_{rt}$. We shall employ a quadratic loss function:

$$
L(\hat{\pi}, \pi) = \sum \limits_{r }  \sum \limits_{t} (\hat{\pi}_{rt} - \pi_{rt})^2.
$$

\begin{figure}[ht]   	
  \begin{center}    
  \minipage{0.49\textwidth}
\centering
\textbf{(a) VOI on the probability scale} 
   \includegraphics[scale=.48, angle = 270]{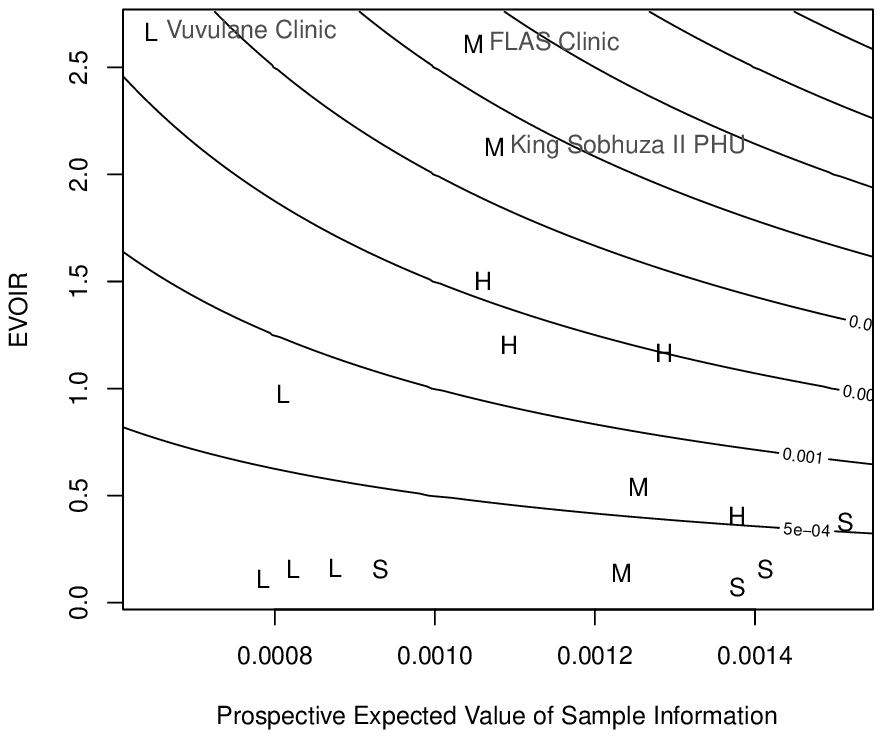} 
   \endminipage\hfill
   \minipage{0.49\textwidth}
\centering
\textbf{(b) VOI on the logit scale} 
    \includegraphics[scale=.48, angle = 270]{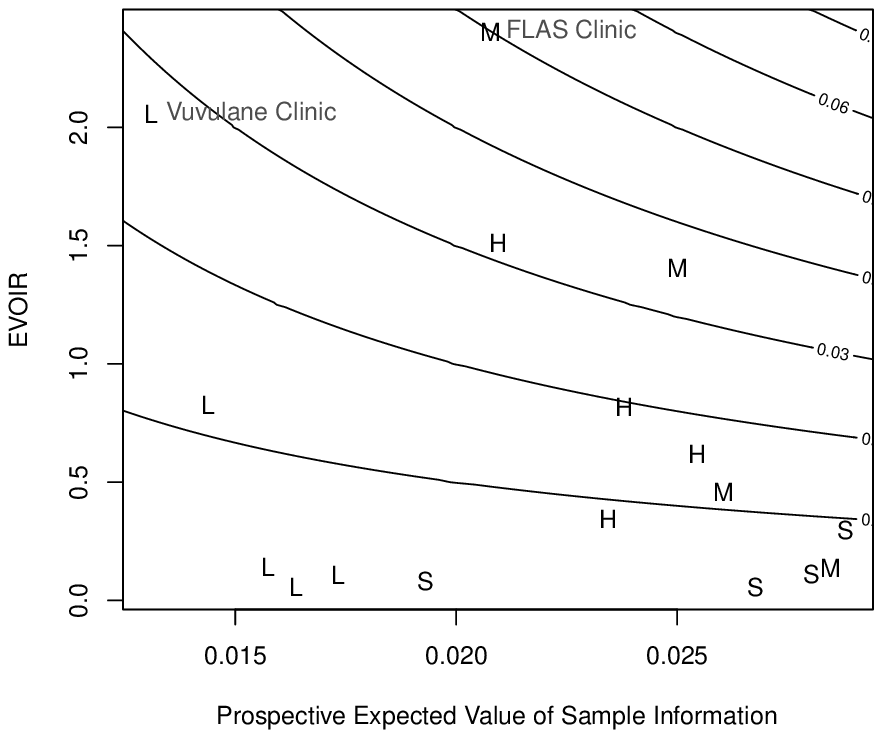}
    \endminipage
    \caption{$\EVOIR$ plotted against $\PVSI$ for each clinic. The plotted letters (L,H,M,S) indicate the region to which the clinic belongs. The three clinics with the highest expected value of information ratios are labeled. The contours indicate $\RVSI$. The left plot (a) uses a squared error loss function on the probability scale, while the right plot (b) uses a quadratic loss function on the logit transformed scale.}
    \label{fig:influence}          
  \end{center}     
\end{figure}

\textbf{Results:} Figure \ref{fig:influence} presents the VOI plots on the probability scale and the logit scale, respectively. The region names are indicated by initial letters (L,H,M,S). Three clinics with the highest $\PVSI$ are all from the Shiselweni region because other regions all have five clinics while Shiselweni only has four. The clinic in Shiselweni has a relatively low $\PVSI$ because of its small sample size of 238, while the average clinic sample size was 695, with the other clinics having sample sizes ranging from 388 to 1026. 
Figure \ref{fig:influence} also presents the full site names for those having high $\EVOIR$ (as twice influential as would have been expected before observing them).
These clinics are indicated as gray dotted lines in Figure S2. One of the clinics, the Vuvulane Clinic, deviates noticeably from the other sites in the Lubombo Region from 2002 to 2006. As indicated by the site's very low $\PVSI$, the data from the Vuvulane Clinic would need to deviate from expectations to a high degree to have a large impact on the model fit. The remaining two sites with a high $\EVOIR$, the FLAS Clinic and King Sobhuza II PHU, are substantially more influential than others. Interestingly, both of these sites are in the Manzini. The numeric values of $\RVSI$, $\PVSI$, and $\EVOIR$ are presented in Supplement 6. 


When choosing which sources of information to investigate for data quality purposes, we typically base the decision on two criteria: how influential the data are and how unusual the data are. If a portion of the data has little to no effect on a decision, any problems with the data will also have little impact on the final decision. On the other hand, data that behaves as expected is unlikely to raise any questions about data quality even if it is influential. Figure \ref{fig:influence} allows us to examine both of these criteria simultaneously. The three clinics labeled by name are likely to be of the highest priority when investigating data quality as the remaining clinics have, at most, a level of influence close to what would be expected ahead of time as indicated by having an $\EVOIR$ close to or less than one. The FLAS clinic, in particular, is simultaneously the most influential and most surprisingly influential of the data sources. Similar conclusions are drawn regardless of whether the original probability scale or the logit transformed scale is used in the squared error loss function.

\section{Discussion}
\label{sec:discussion}

Many existing approaches to Bayesian influence analysis consist of plotting various measures of influence against the index of each observation \citep{kurtek2015,vidal2010,zhu2011}. In this article, we advocate that more insight into the data is granted by considering not just the raw influence of an observation, but also whether this influence is expected based on the rest of the data or if the observation is influential is due to how surprising the observed data are. Our approach to Bayesian influence analysis is to plot all three value of information quantities simultaneously in order to visualize this relationship as illustrated in the U.S. employment rate estimation and the Kingdom of Eswatini HIV prevalence estimation examples. 
By construction, the retrospective expected value of information is the product of the prospective value of information and the expected value of information ratio. Thus we have decomposed the influence of $Y_i$ on the decision into two components: the prospective expected value of information, which measures how far we would have expected the estimate to move by including $Y_i$ had we not observed it, and the expected value of information ratio which measures how much farther the estimate moved than we would have expected. The proposed method was recently used to evaluate the relative contribution of data sources in a multilevel Bayesian hierarchical model for estimating the size of hard-to-reach populations \citep{Parsons2020evaluating}.

There do exist approaches to Bayesian influence analysis that do not reduce to considerations of a single number. These measures tend to focus on identifying ways in which the posterior distribution is changed by including a certain portion of the data. For instance, \citet{weiss1992} suggests evaluating the influence of a single observation by using a plot that includes both the full marginal posterior distribution for a parameter of interest and the marginal posterior computed while excluding the observation. Such a plot allows one to judge any influence the observation has on the posterior distribution of the parameter. \citet{bradlow1997} propose a similar graphical approach in which the influence on all parameters is considered. Such approaches should not be seen as alternatives to our proposal but reasonable next steps. Once observations with high or surprising levels of influence have been identified, graphical approaches can be applied to understand the influence of these observations further.


The primary drawback of the proposed approach lies in the computational difficulty in calculating the VOI quantities. 
There has been recent work on addressing the computational difficulties that arise when attempting to calculate some of the quantities that are central to the value of information methods \citep{ades2004,strong2015,rabideau2018,heath2017,yet2018}. Some level of meta-modeling is generally used in these approaches for computing the expected value of sample information in a reasonable amount of time. 
While exploring all speed-up options is beyond the scope of this paper, we extend the approach of \citet{strong2015} by introducing the $k$ nearest neighbor regression and applying it in Section \ref{sec:swaziland} as a demonstration of the approximation method. 


\section*{Data Availability Statement}
The Longley Data is available as a table in \citet{longley}. The HIV surveillance data in the Kingdom of Eswatini is available at https://aidsinfo.unaids.org/.

\bibliographystyle{abbrvnat}
\bibliography{citations}

\begin{thebibliography}{46}
\providecommand{\natexlab}[1]{#1}
\providecommand{\url}[1]{\texttt{#1}}
\expandafter\ifx\csname urlstyle\endcsname\relax
  \providecommand{\doi}[1]{doi: #1}\else
  \providecommand{\doi}{doi: \begingroup \urlstyle{rm}\Url}\fi

\bibitem[Ades et~al.(2004)Ades, Lu, and Claxton]{ades2004}
A.~E. Ades, G.~Lu, and K.~Claxton.
\newblock {Expected Value of Sample Information Calculations in Medical
  Decision Modeling}.
\newblock \emph{Medical Decision Making}, 24\penalty0 (2):\penalty0 207--227,
  2004.

\bibitem[Ali(1990)]{ali1990}
M.~Ali.
\newblock {A Bayesian Approach to Detect Informative Observations in An
  Experiment}.
\newblock \emph{Communications in Statistics - Theory and Methods}, 19\penalty0
  (7):\penalty0 2567--2575, 1990.

\bibitem[Belsley et~al.(2005)Belsley, Kuh, and Welsch]{Belsley2005regression}
D.~A. Belsley, E.~Kuh, and R.~E. Welsch.
\newblock \emph{Regression diagnostics: Identifying influential data and
  sources of collinearity}, volume 571.
\newblock John Wiley \& Sons, 2005.

\bibitem[Box(1980)]{Box1980sampling}
G.~E. Box.
\newblock Sampling and {B}ayes' inference in scientific modelling and
  robustness.
\newblock \emph{Journal of the Royal Statistical Society: Series A (General)},
  143\penalty0 (4):\penalty0 383--404, 1980.

\bibitem[Bradlow and Zaslavsky(1997)]{bradlow1997}
E.~T. Bradlow and A.~M. Zaslavsky.
\newblock {Case Influence Analysis in Bayesian Inference}.
\newblock \emph{Journal of Computational and Graphical Statistics}, 6\penalty0
  (3):\penalty0 314--331, 1997.

\bibitem[Carlin and Polson(1991)]{carlin1991}
B.~P. Carlin and N.~G. Polson.
\newblock {An Expected Utility Approach to Influence Diagnostics}.
\newblock \emph{Journal of the American Statistical Association}, 86\penalty0
  (416):\penalty0 1013--1021, 1991.
\newblock \doi{10.1080/01621459.1991.10475146}.

\bibitem[Chaloner and Brant(1988)]{Chaloner1988bayesian}
K.~Chaloner and R.~Brant.
\newblock A bayesian approach to outlier detection and residual analysis.
\newblock \emph{Biometrika}, 75\penalty0 (4):\penalty0 651--659, 1988.

\bibitem[Chatterjee and Hadi(2009)]{Chatterjee2009sensitivity}
S.~Chatterjee and A.~S. Hadi.
\newblock \emph{Sensitivity analysis in linear regression}, volume 327.
\newblock John Wiley \& Sons, 2009.

\bibitem[Chatterjee and Yilmaz(1992)]{Chatterjee1992review}
S.~Chatterjee and M.~Yilmaz.
\newblock A review of regression diagnostics for behavioral research.
\newblock \emph{Applied Psychological Measurement}, 16\penalty0 (3):\penalty0
  209--227, 1992.

\bibitem[Cook(1977)]{cook1977}
R.~D. Cook.
\newblock {Detection of Influential Observation in Linear Regression}.
\newblock \emph{Technometrics}, 19\penalty0 (1):\penalty0 15--18, 1977.

\bibitem[Cook and Weisberg(1982)]{Cook1982residuals}
R.~D. Cook and S.~Weisberg.
\newblock \emph{Residuals and influence in regression}.
\newblock New York: Chapman and Hall, 1982.

\bibitem[Geisser(1980)]{Geisser1980discussion}
S.~Geisser.
\newblock Discussion on sampling and bayes' inference in scientific modeling
  and robustness (by g.e.p. box).
\newblock \emph{Journal of the Royal Statistical Society A}, 143:\penalty0
  416--417, 1980.

\bibitem[Geisser(1985)]{Geisser1985prediction}
S.~Geisser.
\newblock On the prediction of observables: a selective update.
\newblock \emph{Bayesian Statistics}, 2:\penalty0 203--230, 1985.

\bibitem[Geisser(1987)]{Geisser1987influential}
S.~Geisser.
\newblock Influential observations, diagnostics and discovery tests.
\newblock \emph{Journal of Applied Statistics}, 14\penalty0 (2):\penalty0
  133--142, 1987.

\bibitem[Gelman and Hill(2006)]{Gelman2006data}
A.~Gelman and J.~Hill.
\newblock \emph{Data analysis using regression and multilevel/hierarchical
  models}.
\newblock Cambridge university press, 2006.

\bibitem[Gelman et~al.(1995)Gelman, Carlin, Stern, and Rubin]{bda}
A.~Gelman, J.~B. Carlin, H.~S. Stern, and D.~B. Rubin.
\newblock \emph{{Bayesian Data Analysis}}.
\newblock Chapman and Hall, 1995.

\bibitem[Gelman et~al.(1996)Gelman, Meng, and Stern]{Gelman1996posterior}
A.~Gelman, X.-L. Meng, and H.~Stern.
\newblock Posterior predictive assessment of model fitness via realized
  discrepancies.
\newblock \emph{Statistica sinica}, pages 733--760, 1996.

\bibitem[Guttman and Pe{\~n}a(1988)]{Guttman1988outliers}
I.~Guttman and D.~Pe{\~n}a.
\newblock Outliers and influence: evaluation by posteriors of parameters in the
  linear model.
\newblock \emph{Bayesian Statistics}, 3:\penalty0 631--640, 1988.

\bibitem[Guttman and Pe{\~n}a(1993)]{Guttman1993bayesian}
I.~Guttman and D.~Pe{\~n}a.
\newblock A {B}ayesian look at diagnostics in the univariate linear model.
\newblock \emph{Statistica Sinica}, pages 367--390, 1993.

\bibitem[Heath et~al.(2017)Heath, Manolopoulou, and Baio]{heath2017}
A.~Heath, I.~Manolopoulou, and G.~Baio.
\newblock {A Review of Methods for Analysis of the Expected Value of
  Information}.
\newblock \emph{Medical Decision Making}, 37\penalty0 (7):\penalty0 747--758,
  2017.

\bibitem[Jackson et~al.(2019)Jackson, Presanis, Conti, and
  De~Angelis]{Jackson2019value}
C.~Jackson, A.~Presanis, S.~Conti, and D.~De~Angelis.
\newblock Value of information: Sensitivity analysis and research design in
  {B}ayesian evidence synthesis.
\newblock \emph{Journal of the American Statistical Association}, 2019.

\bibitem[Johnson and Geisser(1983)]{Johnson1983predictive}
W.~Johnson and S.~Geisser.
\newblock A predictive view of the detection and characterization of
  influential observations in regression analysis.
\newblock \emph{Journal of the American Statistical Association}, 78\penalty0
  (381):\penalty0 137--144, 1983.

\bibitem[Johnson and Geisser(1985)]{Johnson1985estimative}
W.~Johnson and S.~Geisser.
\newblock Estimative influence measures for the multivariate general linear
  model.
\newblock \emph{Journal of Statistical Planning and Inference}, 11\penalty0
  (1):\penalty0 33--56, 1985.

\bibitem[Keisler et~al.(2014)Keisler, Collier, Chu, Sinatra, and
  Linkov]{keisler2014}
J.~M. Keisler, Z.~A. Collier, E.~Chu, N.~Sinatra, and I.~Linkov.
\newblock {Value of Information Analysis: the State of Application}.
\newblock \emph{Environment Systems and Decisions}, 34\penalty0 (1):\penalty0
  3--23, Mar 2014.
\newblock ISSN 2194-5411.

\bibitem[Kempthorne(1986)]{kempthorne1986}
P.~J. Kempthorne.
\newblock {Decision-Theoretic Measures of Influence in Regression}.
\newblock \emph{Journal of the Royal Statistical Society: Series B
  (Methodological)}, 48\penalty0 (3):\penalty0 370--378, 1986.

\bibitem[Kurtek and Bharath(2015)]{kurtek2015}
S.~Kurtek and K.~Bharath.
\newblock {Bayesian Sensitivity Analysis with the Fisher--Rao Metric}.
\newblock \emph{Biometrika}, 102\penalty0 (3):\penalty0 601--616, July 2015.

\bibitem[Lindley(1956)]{lindley1956}
D.~V. Lindley.
\newblock {On a Measure of the Information Provided by an Experiment}.
\newblock \emph{The Annals of Mathematical Statistics}, 27\penalty0
  (4):\penalty0 986--1005, 1956.

\bibitem[Longley(1967)]{longley}
J.~W. Longley.
\newblock {An Appraisal of Least Squares Programs for the Electronic Computer
  from the Point of View of the User}.
\newblock \emph{Journal of the American Statistical Association}, 62\penalty0
  (319):\penalty0 819--841, 1967.
\newblock ISSN 01621459.

\bibitem[Mahalanobis(1936)]{Mahalanobis1936}
P.~C. Mahalanobis.
\newblock On the generalized distance in statistics.
\newblock In \emph{National Institute of Science of India}, 1936.

\bibitem[McCulloch(1989)]{Mcculloch1989local}
R.~E. McCulloch.
\newblock Local model influence.
\newblock \emph{Journal of the American Statistical Association}, 84\penalty0
  (406):\penalty0 473--478, 1989.

\bibitem[Millar et~al.(2007)Millar, Stewart, et~al.]{Millar2007assessment}
R.~B. Millar, W.~S. Stewart, et~al.
\newblock Assessment of locally influential observations in {B}ayesian models.
\newblock \emph{Bayesian Analysis}, 2\penalty0 (2):\penalty0 365--383, 2007.

\bibitem[Parsons et~al.(2020)Parsons, Niu, and Bao]{Parsons2020evaluating}
J.~Parsons, X.~Niu, and L.~Bao.
\newblock Evaluating the relative contribution of data sources in a {B}ayesian
  analysis with the application of estimating the size of hard to reach
  populations.
\newblock \emph{Statistical Communications in Infectious Diseases}, 12\penalty0
  (s1), 2020.

\bibitem[Pe{\~n}a and Guttman(1993)]{Pena1993comparing}
D.~Pe{\~n}a and I.~Guttman.
\newblock Comparing probabilistic methods for outlier detection in linear
  models.
\newblock \emph{Biometrika}, 80\penalty0 (3):\penalty0 603--610, 1993.

\bibitem[Peng and Dey(1995)]{Peng1995bayesian}
F.~Peng and D.~K. Dey.
\newblock Bayesian analysis of outlier problems using divergence measures.
\newblock \emph{Canadian Journal of Statistics}, 23\penalty0 (2):\penalty0
  199--213, 1995.

\bibitem[Rabideau et~al.(2018)Rabideau, Pei, Walensky, Zheng, and
  Parker]{rabideau2018}
D.~J. Rabideau, P.~P. Pei, R.~P. Walensky, A.~Zheng, and R.~A. Parker.
\newblock {Implementing Generalized Additive Models to Estimate the Expected
  Value of Sample Information in a Microsimulation Model: Results of Three Case
  Studies}.
\newblock \emph{Medical Decision Making}, 38\penalty0 (2):\penalty0 189--199,
  2018.

\bibitem[Raiffa and Schlaifer(1961)]{raiffa1961}
H.~Raiffa and R.~Schlaifer.
\newblock \emph{{Applied Statistical Decision Theory}}.
\newblock {Studies in Managerial Economics}. Division of Research, Graduate
  School of Business Adminitration, Harvard University, 1961.

\bibitem[Smith and Pettit(1985)]{pettit1985}
A.~F.~M. Smith and L.~I. Pettit.
\newblock {Outliers and Influential Observations in Linear Models}.
\newblock \emph{Bayesian Statistics}, 2:\penalty0 473--494, 1985.

\bibitem[Strong et~al.(2015)Strong, Oakley, Brennan, and Breeze]{strong2015}
M.~Strong, J.~E. Oakley, A.~Brennan, and P.~Breeze.
\newblock {Estimating the Expected Value of Sample Information Using the
  Probabilistic Sensitivity Analysis Sample: A Fast, Nonparametric
  Regression-Based Method }.
\newblock \emph{Medical Decision Making}, pages 570--583, July 2015.

\bibitem[Van Der~Linde et~al.(2007)]{Van2007local}
A.~Van Der~Linde et~al.
\newblock Local influence on posterior distributions under multiplicative modes
  of perturbation.
\newblock \emph{Bayesian Analysis}, 2\penalty0 (2):\penalty0 319--332, 2007.

\bibitem[Vidal and Castro(2010)]{vidal2010}
I.~Vidal and L.~M. Castro.
\newblock {Influential Observations in the Independent Student- t Measurement
  Error Model with Weak Nondifferential Error}.
\newblock \emph{Chilean Journal of Statistics}, 1\penalty0 (2):\penalty0
  17--34, September 2010.

\bibitem[Weiss(1996)]{weiss1996}
R.~Weiss.
\newblock {An Approach to Bayesian Sensitivity Analysis}.
\newblock \emph{Journal of the Royal Statistical Society: Series B
  (Methodological)}, 58\penalty0 (4):\penalty0 739--750, 1996.

\bibitem[Weiss and Cook(1992)]{weiss1992}
R.~E. Weiss and R.~D. Cook.
\newblock {A Graphical Case Statistic for Assessing Posterior Influence}.
\newblock \emph{Biometrika}, 79\penalty0 (1):\penalty0 51--55, March 1992.

\bibitem[Yet et~al.(2018)Yet, Constantinou, Fenton, and Neil]{yet2018}
B.~Yet, A.~Constantinou, N.~Fenton, and M.~Neil.
\newblock {Expected Value of Partial Perfect Information in Hybrid Models Using
  Dynamic Discretization}.
\newblock \emph{IEEE Access}, 6:\penalty0 7802--7817, 2018.

\bibitem[Zellner(1975)]{Zellner1975bayesian}
A.~Zellner.
\newblock Bayesian analysis of regression error terms.
\newblock \emph{Journal of the American Statistical Association}, 70\penalty0
  (349):\penalty0 138--144, 1975.

\bibitem[Zellner(1986)]{Zellner1986assessing}
A.~Zellner.
\newblock On assessing prior distributions and {B}ayesian regression analysis
  with g-prior distributions.
\newblock \emph{Bayesian Inference and Decision Techniques}, 1986.

\bibitem[Zhu et~al.(2011)Zhu, Ibrahim, and Tang]{zhu2011}
H.~Zhu, J.~G. Ibrahim, and N.~Tang.
\newblock {Bayesian Influence Analysis: A Geometric Approach}.
\newblock \emph{Biometrika}, 98\penalty0 (2):\penalty0 307--323, June 2011.

\end{thebibliography}

\end{document}


\def\spacingset#1{\renewcommand{\baselinestretch}%
{#1}\small\normalsize} \spacingset{1.3}


\if0\blind
{
\title{Supplementary Materials for\\
``A Unified Approach for Outliers and Influential Data Detection  -- The Value of Information in Retrospect"}
  \maketitle
} \fi

\if1\blind
{
  \bigskip
  \bigskip
  \bigskip
  \begin{center}
    {\LARGE\bf Supplementary Materials for\\
``A Unified Approach for Outliers and Influential Data Detection  -- The Value of Information in Retrospect"}
\end{center}
  \medskip
} \fi

\noindent
S1. Properties of Proposed Measures Under Quadratic Loss Function.\\
S2. Properties of Proposed Measures in Linear Regression With Non-Informative Prior.\\
S3. Properties of Proposed Measures in Linear Regression With $g$-Prior.\\
S4. Approximation of $E\{\pi \mid Y^{(k)}\}$ for Eswatini Example.\\
S5. Table of Value of Information Measures for Longley Data.\\
S6. HIV Prevalence Estimates and VOI Measures for Eswatini Example.
\par

\section{Properties of Proposed Measures Under Quadratic Loss Function}

Let $Y_{-i}$ and $Y_i$ be random vectors of finite dimension. Let $\theta$ be a parameter that we are interested in estimating. We shall use a quadratic loss function, $L(a, \theta) = (a - \theta) ^ T Q (a - \theta)$, to measure how good an estimate $a$ is, where $Q = A^T A \in \mathbb{R}^{p \times p}$ is a positive definite matrix. We shall assume that the expectation of $\theta$ conditional on $Y_{-i}$ and the expectation conditional on  $Y_{-i}$ and $Y_i$ both exist and are finite. It is easy to show that the Bayes action in each case is the conditional mean of the parameter $\theta$,
\begin{align*}
  a_{Y_{-i}} &= E \big ( \theta \mid Y_{-i} \big ), & a_{Y_{-i}, Y_i} &= E \big ( \theta \mid Y_{-i}, Y_i \big ).
\end{align*}

We shall make use of the fact that $E[a_{Y_{-i}, Y_i} \mid Y_{-i}] = a_{Y_{-i}}$, a consequence of the law of total expectation. We may establish the expression for the retrospective expected value of sample information given by Theorem 1 as follows:
$$
\begin{aligned}
  E \big \{ \PVSI(Y_i \mid Y_{-i}; \theta) \mid Y_{-i},Y_i \big \} &= E \big \{ L(a_{Y_{-i}}, \theta) - L(a_{Y_{-i},Y_i}, \theta)  \mid Y_{-i},Y_i  \big \}  \\
  &= E \big \{ (a_{Y_{-i}} - \theta)^TQ(a_{Y_{-i}}-\theta) - (a_{Y_{-i}, Y_i} - \theta)^TQ(a_{Y_{-i}, Y_i}-\theta)   \mid Y_{-i},Y_i  \big \} \\
  &= E \big ( a_{Y_{-i}}^T Q a_{Y_{-i}} - 2 a_{Y_{-i}}^T Q \theta -  a_{Y_{-i}, Y_i}^T Q a_{Y_{-i}, Y_i} + 2 a_{Y_{-i}, Y_i}^T Q \theta  \mid Y_{-i},Y_i  \big ) \\ 
  &= a_{Y_{-i}}^T Q a_{Y_{-i}} - 2 a_{Y_{-i}}^T Q  a_{Y_{-i}, Y_i} -  a_{Y_{-i}, Y_i}^T Q a_{Y_{-i}, Y_i} + 2 a_{Y_{-i}, Y_i}^T Q  a_{Y_{-i}, Y_i} \\
  &= a_{Y_{-i}}^T Q a_{Y_{-i}} - 2 a_{Y_{-i}}^T Q  a_{Y_{-i}, Y_i} + a_{Y_{-i}, Y_i}^T Q  a_{Y_{-i}, Y_i} \\
  &= (a_{Y_{-i}} - a_{Y_{-i},Y_i})^T Q (a_{Y_{-i}} - a_{Y_{-i},Y_i}).
\end{aligned}
$$
The expression for the prospective expected value of sample information described by Theorem 2 can be demonstrated as follows:
$$
\begin{aligned}
  E \big \{ \PVSI(Y_i \mid Y_{-i}; \theta) \mid Y_{-i} \big \} &= E \Big [ E \big \{ \PVSI(Y_i \mid Y_{-i}; \theta)  \mid Y_{-i},Y_i \big \}  \mid Y_{-i} \Big ] \\
  &= E \big \{ (a_{Y_{-i}} - a_{Y_{-i},Y_i})^T Q (a_{Y_{-i}} - a_{Y_{-i},Y_i})  \mid Y_{-i} \big \} \\
  &= E \big \{  (A a_{Y_{-i},Y_i} - A a_{Y_{-i}} )^T (A a_{Y_{-i},Y_i} - A a_{Y_{-i}} )   \mid Y_{-i} \big \} \\
  &= tr\Big \{ var(A  a_{Y_{-i},Y_i} \mid Y_{-i}  ) \Big\}.
\end{aligned}
$$
Next we demonstrate Theorem 3. Suppose that $a_{Y_{-i},Y_i} \mid Y_{-i} \sim N(a_{Y_{-i}}, c\Sigma)$ such that $c > 0$ and $\Sigma Q$ is idempotent.  Then, the moment generating function for $p \EVOIR(Y_i \mid Y_{-i})$ for $t<\delta$ for some $\delta > 0$ is given by
$$
\begin{aligned}
m_E (t) &=  E[ \exp \Big( t  \times p \EVOIR(Y_i \mid Y_{-i}) \Big) \vert Y_{-i} ] \\
&=  E \Big[ \exp \Big( t \times p \EVOIR(Y_i \mid Y_{-i}) \Big) \vert Y_{-i}\Big] \\
&=  E \Big[ \exp\Big( t \times p \frac{(a_{Y_{-i}} - a_{Y_{-i},Y_i})^T Q (a_{Y_{-i}} - a_{Y_{-i},Y_i}) }{tr\Big \{ var(A  a_{Y_{-i},Y_i} \mid Y_{-i}  ) \Big\}}\Big) \Big \vert Y_{-i}  \Big]\\
&= \int \exp\Big( t \times p \frac{(x - a_{Y_{-i}})^T Q (x - a_{Y_{-i}}) }{tr\Big (c A \Sigma A^T \Big) }\Big) \\
& \quad \quad  \quad  \quad (2 \pi)^{-k/2} det(c \Sigma)^{-1/2} \exp\Big( - \frac{1}{2} (x - a_{Y_{-i}})^T (c\Sigma)^{-1} (x - a_{Y_{-i}})  \Big) \,dx \\
&= \int  (2 \pi)^{-k/2} det(c \Sigma)^{-1/2} \exp\Big( - \frac{1}{2} (x - a_{Y_{-i}})^T \Big( c c^{-1}\Sigma^{-1} - \frac{2tp}{tr(c A \Sigma A^T)} Q \Big)(x - a_{Y_{-i}})  \Big) \,dx \\
&= det\Big(c \Sigma\Big)^{-1/2} det \Big(c^{-1}\Sigma^{-1} - \frac{2tp}{tr(c A \Sigma A^T)} Q \Big)^{-1/2} \\
&=  det\Big( I_k  - \frac{2tp}{tr(A \Sigma A^T)} \Sigma  Q\Big)^{-1/2} 
\end{aligned}
$$
However, due to the cyclic property of the trace of a matrix,
$$
tr(A \Sigma A^T) = tr(\Sigma A^T A) = tr(\Sigma Q).
$$
However, since $\Sigma Q$ is idempotent and $Q$ is full rank,
$$
tr(A \Sigma A^T) = rank(\Sigma Q) = p.
$$
So,
$$
\begin{aligned}
m_E (t) &= det\Big( I_k  - 2t \Sigma Q\Big)^{-1/2} .\\
\end{aligned}
$$
Since $\Sigma Q$ is idempotent, it is also diagonalizable and can be written as
$$
\Sigma Q = P D P^{-1}
$$
where $P$ is is invertible and $D$ is a diagonal matrix whose entries consist of exactly $p$ values equal to $1$ and the remaining values equal to $0$. So,
\[
\begin{aligned}
m_E (t) &= det\Big( I_k  - 2t  P D P^{-1}  \Big)^{-1/2} \\
& = det\Big(  P ( I_k - 2tD) P^{-1} \Big)^{-1/2}\\
&= det\Big(  I_k - 2tD \Big)^{-1/2}\\
&= (1 - 2t)^{-p/2}
\end{aligned}
\]
This is the moment generating function for a chi-square distribution having $p$ degrees of freedom, proving Theorem 3. 


\section{Properties of Proposed Measures in Linear Regression With Non-Informative Prior}

The retrospective expected value of sample information in Corollary 1 may be obtained from Theorem 1 as follows:
$$
\begin{aligned}
  E \big \{ \VSI(Y_i \mid Y_{-i}; \beta)  \mid Y \big \} &= (\hat{\beta} - \hat{\beta}_{-i})^T X^TX (\hat{\beta} - \hat{\beta}_{-i}) \\
  &= (X\hat{\beta} - X\hat{\beta}_{-i})^T(X\hat{\beta} - X\hat{\beta}_{-i}) \\
  &= \sum_{k = 1}^n (X_k \hat{\beta} - X_k \hat{\beta}_{-i})^2
\end{aligned} 
$$

Here we derive the form of the prospective expected value of sample information in Corollary 2. We begin by applying Theorem 2:
$$
\begin{aligned}
  E \big \{ \PVSI(Y_i \mid Y_{-i}; \theta)  \mid Y_{-i} \big \} &= tr\Big\{X var \big ( \hat{\beta}_i \mid Y_{-i} \big) X^T\Big\} \\
  &= \sum_{k = 1}^n X_k var \big ( \hat{\beta}_i \mid Y_{-i} \big) X^T_k  \\
  &= \sum_{k = 1}^n X_k var \Big \{ (X^T X)^{-1}X^T Y \mid Y_{-i} \Big \} X^T_k  \\
  &=  \sum_{k = 1}^n X_k (X^T X)^{-1}X^T var (Y \mid Y_{-i}) X (X^T X)^{-1}X_k^T \\
  &=  \sum_{k = 1}^n X_k (X^T X)^{-1}X_i^T var (Y_i \mid Y_{-i}) X_i (X^T X)^{-1}X_k^T \\
  &= \sum_{k = 1}^n h^2_{ik}  var (Y_i \mid Y_{-i})\\
  &= h_{ii}  var (Y_i \mid Y_{-i}).
\end{aligned}
$$
The last equality follows from the idempotence of the hat matrix $H$. The predictive variance for the $i$th observation is
$$
\begin{aligned}
  var(Y_i \mid Y_{-i}) &= \frac{n-p-1}{n-p-3} S_{-i}^2 \{1 + X_i(X^T_{-i}X_{-i})^{-1}X_i^T\}.
\end{aligned}
$$
So,
$$
\begin{aligned}
  E \big \{ \PVSI(Y_i \mid Y_{-i}; \theta)  \mid Y_{-i} \big \}  &= h_{ii} \frac{n-p-1}{n-p-3} S_{-i}^2 \{1 + X_i(X^T_{-i}X_{-i})^{-1}X_i^T\} .
\end{aligned}
$$
An application of the Sherman--Morrison--Woodbury formula gives
$$
\begin{aligned}
  (X_{-i}^T X_{-i})^{-1} &= (X^TX - X_i^T X_i)^{-1} \\
  &= (X^TX)^{-1} + \frac{(X^TX)^{-1} X_i^T X_i (X^TX)^{-1}}{1 - X_i(X^TX)^{-1} X_i^T}.
\end{aligned}
$$

Thus,
$$
\begin{aligned}
  X_i(X^T_{-i}X_{-i})^{-1}X_i^T &=  X_i(X^TX)^{-1}X_i^T + \frac{X_i(X^TX)^{-1} X_i^T X_i (X^TX)^{-1}X_i^T}{1 - X_i(X^TX)^{-1} X_i^T}\\
  &= h_{ii} +  \frac{ h_{ii}^2}{1 -  h_{ii}}\\
  &=  \frac{ h_{ii}}{1 -  h_{ii}}.
\end{aligned}
$$
So, the form of the prospective expected value of sample information is,
$$
\begin{aligned}
  E \big \{ \PVSI(Y_i \mid Y_{-i}; \theta)  \mid Y_{-i} \big \} & = h_{ii} \frac{n-p-1}{n-p-3} S_{-i}^2 \{1 + X_i(X^T_{-i}X_{-i})^{-1}X_i^T\} \\
  & = h_{ii} \frac{n-p-1}{n-p-3} S_{-i}^2 \frac{1}{1 - h_{ii}}                      \\
  & = \frac{n-p-1}{n-p-3} S_{-i}^2 \frac{h_{ii}}{1 - h_{ii}}.
\end{aligned}
$$

Finally, under the assumptions of Corollary 1, the expected value of information ratio in Corollary 3 is
$$
\begin{aligned}
  \EVOIR(Y_i \mid Y_{-i}) &= \frac{ \sum_{k = 1}^n (X_k\hat{\beta} - X_k\hat{\beta}_{-i})^2}{\frac{n-p-1}{n-p-3} S_{-i}^2 \frac{h_{ii}}{1 - h_{ii}}}\\
  &=  \frac{(n-p-3) \frac{h_{ii}}{(1 - h_{ii})^2} (Y_i - X_i\hat{\beta})^2 }{(n-p-1) S_{-i}^2 \frac{h_{ii}}{1 - h_{ii}}}\\
  &= \frac{(n-p-3)(Y_i - X_i\hat{\beta})^2 }{(n-p-1) S_{-i}^2 (1 - h_{ii})}\\
  &=  \frac{(n-p-3) }{(n-p-1)}t_{(i)}^2.
\end{aligned}
$$
Here $t_{(i)}$ is the externally Studentized residual for the $i$th observation. Note that the predictive distribution of $Y_i$ is
$$
\frac{Y_i - X_i \hat{\beta}_{-i}}{ S_{-i}  \sqrt{ (1 - h_{ii})^{-1} }}  \Big\vert Y_{-i} \sim  t_{n-p-1} .
$$
It follows that
$$
\begin{aligned}
  \EVOIR(Y_i \mid Y_{-i}) \sim \frac{(n-p-3) }{(n-p-1)} F_{1, n-p-1}.
\end{aligned}
$$
\\

\section{Properties of Proposed Measures in Linear Regression With g-Prior}

Next, we consider the informative $g$-prior distribution for $\beta$ \citep{Zellner1986assessing} and the conjugate prior distribution for $\sigma^2$:
$$
\begin{aligned}
\beta &\sim \textup{normal}(\beta_0, g \sigma^2 (X^TX)^{-1}),\\
\sigma^2 &\sim \textup{inverse-gamma}(\nu_0/2, \nu_0\sigma_0^2/2).
\end{aligned}
$$ 

We center the response and predictors so that the intercept is omitted and set $\beta_0 = 0$ which corresponds to no associations between the response and the covariates in the prior distribution. The prior distribution of $\sigma^2$ is weakly centered around the least square estimate $\sigma_{\textup{ls}}^2$ by taking $\nu_0=1$ and $\sigma_0^2 = \sigma_{\textup{ls}}^2$. It can be shown that $p(\beta|Y,X,\sigma^2)$ is a multivariate normal distribution:
$$
\beta|Y,X,\sigma^2 \sim \textup{normal}(\frac{g}{g+1} \hat{\beta}_{\textup{ls}}, \frac{g}{g+1}  \sigma^2 (X^TX)^{-1}).
$$
See the proof in \cite{Hoff2009first} Section 9.2.2. The Bayes action of estimating $\beta$ is the posterior mean, $\frac{g}{g+1} \hat{\beta}_{\textup{ls}}$, which does not depend on $\sigma^2$. In the appendix, it is shown that the Bayes action when estimating $\beta$ conditional only on $Y_{-i}$ is
$$
g(X^TX + gX^T_{-i}X_{-i} )^{-1} X^T_{-i} Y_{-i} 
$$
Notice that this estimate converges to the least squares estimate based on the reduced data as the prior becomes weaker, i.e. $g\to \infty$. This fact can be seen by rewriting the estimate under the reduced data in the following way:
$$
\frac{g}{g+1} \Big(I_p - \frac{1}{1 + g + X_i (X^T_{-i} X_{-i}) X_i^T } (X^T_{-i} X_{-i})^{-1} X^T_i X_i \Big) \hat{\beta}_{-i} .
$$
The retrospective value of sample information, $\RVSI$, has a very similar form to that for the non-informative prior:
\begin{equation}
\label{eqn:gRVOI}
\frac{g^2}{(g+1)^2} D_g(\hat{\beta}_{ls}, \hat{\beta}_{-i} ) ^T X^TX D_g(\hat{\beta}_{ls}, \hat{\beta}_{-i} ), 
\end{equation}
whose limit is the non-informative retrospective value as the prior becomes less informative ($g\to \infty$) and is zero when $g=0$, where
$$
D_g(\hat{\beta}_{ls}, \hat{\beta}_{-i} ) = \hat{\beta}_{ls} - \hat{\beta}_{-i} + \frac{1}{1 + g + X_i (X^T_{-i} X_{-i}) X_i^T } (X^T_{-i} X_{-i})^{-1} X^T_i X_i \hat{\beta}_{-i} .
$$
The prospective value of sample information, $\PVSI$,  becomes
\begin{equation}
\label{eqn:gPVOI}
\frac{g^2}{(g+1)^2} \hat{\sigma}_{-i,g}^2 h_{ii}\Big(  \frac{1 + g}{1 + g(1 - h_{ii}) }    \Big),
\end{equation}
where
$$
\hat{\sigma}_{-i,g}^2 = E[ \sigma \vert Y_{-i} ] = \frac{ \sigma^2_{ls} + Y_{-i}^T[I_{n-1} + gX_{-i} (X^TX)^{-1} X_{-i}^T  ]^{-1} Y_{-i}  }{n - 2 }.
$$
This can be rewritten using the Woodbury matrix identity to
$$
\hat{\sigma}_{-i,g}^2 =  \frac{\sigma^2_{ls} + Y_{-i}^T\big(I_{n-1} - gX_{-i} [X^TX + g X^T_{-i} X_{-i}]^{-1} X^T_{-i} \big) Y_{-i} }{n - 2 }.
$$
The prospective value of sample information, $\EVOIR$, tends towards the a modified version non-informative case when the prior becomes weaker and the modification only involves using a different estimate for the variance parameter. 
 
Finally, the expected value of information ratio is, as before, the ratio of these two quantities, and it has a non-zero limit as $g\to 0$:
\begin{equation}
\label{eqn:gEVOIRlimit}
\frac{D_0(\hat{\beta}_{ls}, \hat{\beta}_{-i} ) ^T X^TX D_0(\hat{\beta}_{ls}, \hat{\beta}_{-i} ) }{\hat{\sigma}^2_{-i,0} h_{ii}}.
\end{equation}

We show that
\begin{equation}
\label{eqn:gEVOIR}
\EVOIR(Y_i \mid Y_{-i}) = \frac{n-2}{n} T_i^{2}
\sim \frac{n-2}{n} F_{1,n},
\end{equation}
where $T_i \vert Y_{-i} \sim t_n$. \\

\textbf{Derivation:} In what follows we will present the detailed derivations of $\RVSI$, $\PVSI$ and $\EVOIR$ under the $g$-prior and assume that the response and predictors are centered.

Since $Y = X \beta + \epsilon$ where $\beta$ and $\epsilon = Y - X\beta$ are independent conditional on $\sigma^2$,
$$
Y \vert \sigma^2 \sim N\Big(\mathbf{0}, \sigma^2 [ I_n + g X(X^TX)^{-1}X^T ] \Big ).
$$
As the inverse gamma distribution is conjugate for the normal distribution,
$$
\sigma^2 \vert Y \sim \text{ Inverse-Gamma}\Big(\frac{n+1}{2} , \frac{\sigma_0^2 + Y^T [ I_n + g X(X^TX)^{-1}X^T ]^{-1} Y }{2}\Big)
$$
and
$$
\sigma^2 \vert Y_{-i} \sim \text{ Inverse-Gamma}\Big(\frac{n}{2} , \frac{\sigma_0^2 + Y_{-i}^T [ I_{n-1} + g X_{-i}(X^TX)^{-1}X_{-i}^T ]^{-1} Y_{-i} }{2} \Big)
$$
The posterior mean of $\sigma^2$ is therefore
$$
E[\sigma^2 \vert Y] = \frac{\sigma_0^2 + Y^T [ I_n + g X(X^TX)^{-1}X^T ]^{-1} Y }{ n - 1}\Big).
$$
Similarly, the posterior mean conditional on only the reduced data is
$$
E[\sigma^2 \vert Y_{-i}] = \frac{\sigma_0^2 + Y_{-i}^T [ I_{n-1} + g X_{-i}(X^TX)^{-1}X_{-i}^T ]^{-1} Y_{-i}  }{ n - 2}\Big).
$$
Let $\hat{\beta}_{-i} =(X_{-i}^T X_{-i})^{-1} X^T_{-i} Y_{-i}$. Then, 
$$
\hat{\beta}_{-i} \vert \beta, \sigma^2 \sim N\Big(\beta, \sigma [X^T_{-i} X_{-i}]^{-1} \Big)
$$
Using the fact that a multivariate normal distribution is conjugate for the mean of multivariate normal with a known covariance matrix we arrive at
$$
\beta \vert \hat{\beta}_{-i}, \sigma^2 \sim N( \mu_i, \sigma^2 \Sigma_i)
$$
where
$$
\mu_i = g [X^TX + gX^T_{-i} X_{-i}]^{-1} X^T_{-i}X_{-i} \hat{\beta}_{-i}
$$
and
$$
\Sigma_i = g [X^TX + gX^T_{-i} X_{-i}]^{-1} .
$$
Noting that $\hat{\beta}_{-i}$ is a sufficient statistic of $Y_{-i}$ for $\beta$ given fixed $\sigma^2$, we see that 
$$
\beta \vert Y_{-i}, \sigma^2 \sim N( \mu_i, \sigma^2 \Sigma_i).
$$
An application of the fact that $X^TX = X_{-i}^TX_{-i} + X^T_iX_i$ and then the Woodbury matrix identity gives
$$
\begin{aligned}
g[X^TX + gX^T_{-i} X_{-i}]^{-1} &= g [X_{-i}^TX_{-i} + X^T_iX_i + gX^T_{-i} X_{-i}]^{-1} \\
&= g [ (1+g)X^T_{-i} X_{-i} + X^T_iX_i]^{-1} \\
&= g \Big[ \frac{1}{1+g}(X^T_{-i} X_{-i} )^{-1} - \Big( \frac{1}{(1+g)^2} \Big) \Big( \frac{(X^T_{-i} X_{-i} )^{-1}  X^T_iX_i (X^T_{-i} X_{-i} )^{-1}}{1+(1+g)^{-1} X_i (X^T_{-i} X_{-i} )^{-1} X^t_i} \Big)  \Big] \\
&= \frac{g}{g+1} \Big[ (X^T_{-i} X_{-i} )^{-1} -  \frac{(X^T_{-i} X_{-i} )^{-1}  X^T_iX_i (X^T_{-i} X_{-i} )^{-1}}{1+g + X_i (X^T_{-i} X_{-i} )^{-1} X^t_i}  \Big]
\end{aligned}
$$
As $\mu_i$ does not depend on $\sigma^2$, we see that
$$
\begin{aligned}
E[\beta \vert Y_{-i}] &=  g [X^TX + gX^T_{-i} X_{-i}]^{-1} X^T_{-i}X_{-i} \hat{\beta}_{-i}\\
&= \frac{g}{g+1} \Big[ (X^T_{-i} X_{-i} )^{-1} -  \frac{(X^T_{-i} X_{-i} )^{-1}  X^T_iX_i (X^T_{-i} X_{-i} )^{-1}}{1+g + X_i (X^T_{-i} X_{-i} )^{-1} X^t_i}  \Big] X^T_{-i} X_{-i} \hat{\beta}_{-i} \\
&= \frac{g}{g+1} \Big[ \hat{\beta}_{-i}-  \frac{1}{1+g + X_i (X^T_{-i} X_{-i} )^{-1} X^t_i}  (X^T_{-i} X_{-i} )^{-1}  X^T_iX_i \hat{\beta}_{-i} \Big] .
\end{aligned}
$$
Recall that the posterior mean of $\beta$ conditional on the complete data is
$$
E[\beta \vert Y] = \frac{g}{1+g} \hat{\beta}_{ls} .
$$
From this, we can apply Theorem 1 to obtain the retrospective value of sample information:
$$
 E \big \{ \PVSI(Y_i \mid Y_{-i}; \theta) \mid Y \big \} = \frac{g^2}{(g+1)^2} D_g(\hat{\beta}_{ls}, \hat{\beta}_{-i} ) ^T X^TX D_g(\hat{\beta}_{ls}, \hat{\beta}_{-i} ) 
$$
where, as in the main text,
$$
D_g(\hat{\beta}_{ls}, \hat{\beta}_{-i} ) = \hat{\beta}_{ls} - \hat{\beta}_{-i} + \frac{1}{1 + g + X_i (X^T_{-i} X_{-i}) X_i^T } (X^T_{-i} X_{-i})^{-1} X^T_i X_i \hat{\beta}_{-i} .
$$
The formula for the prospective expected value of sample information is derived in a similar manner to the non-informative case using Theorem 2:
$$
\begin{aligned}
 E \big \{ \PVSI(Y_i \mid Y_{-i}; \theta) \mid Y_{-i} \big \} &= tr \Big \{  X Var \big ( \frac{g}{1+g} \hat{\beta}_{ls} \vert Y_{-i} \big ) X^T \Big \} \\
 &= tr \Big \{  \frac{g^2}{(1+g)^2}  X Var \big (  \hat{\beta}_{ls} \vert Y_{-i} \big ) X^T \Big \} \\
 &= \sum \limits_{k=1}^{n} \frac{g^2}{(1+g)^2} X_k Var \big (  \hat{\beta}_{ls} \vert Y_{-i} \big ) X_k^T\\
 &=  \sum \limits_{k=1}^{n} \frac{g^2}{(1+g)^2} X_k (X^TX)^{-1} X^T Var \big (  Y \vert Y_{-i} \big ) X (X^TX)^{-1}   X_k^T \\
 &= \sum \limits_{k=1}^{n} \frac{g^2}{(1+g)^2} X_k (X^TX)^{-1} X_i^T Var \big (  Y_i \vert Y_{-i} \big ) X_i (X^TX)^{-1}   X_k^T \\
 &= \sum \limits_{k=1}^{n} \frac{g^2}{(1+g)^2} h_{ik}^2 Var \big (  Y_i \vert Y_{-i} \big ) \\
 &= \frac{g^2}{(1+g)^2} h_{ii} Var(Y_i \vert Y_{-i}) .
\end{aligned}
$$
However, the pair $(Y_i, \sigma^2)$ has a normal inverse gamma distribution conditional on $Y_{-i}$:
$$
(Y_i, \sigma^2) \vert Y_{-i} \sim NIG\Big(X_i \mu_i, (1 + X_i \Sigma_i X_i^T)^{-1}, \frac{n}{2} , b_i \Big)
$$
where
$$
b_i = \frac{\sigma_0^2 + Y_{-i}^T [ I_{n-1} + g X_{-i}(X^TX)^{-1}X_{-i}^T ]^{-1} Y_{-i}}{2} .
$$
It follows that
$$
\sqrt{\frac{n(1 + X_i \Sigma_i X_i^T)^{-1}}{2 b_i} } \Big( Y_i - X_i \mu_i \Big) \vert Y_{-i} \sim t_n.
$$
From this we can conclude that
$$
\begin{aligned}
 Var(Y_i \vert Y_{-i}) &= \frac{2 b_i}{n(1 + X_i \Sigma_i X_i^T)^{-1}} \frac{n}{n - 2} \\
 &= \frac{2 b_i}{n - 2} (1 + X_i \Sigma_i X_i^T)\\
 &= E[\sigma^2 \vert Y_{-i} ](1 + g X_i [X^TX + gX^T_{-i} X_{-i}]^{-1} X_i^T).
\end{aligned}
$$
This expression can be further simplified using the Sherman–Morrison formula:
$$
\begin{aligned}
g X_i [X^TX + gX^T_{-i} X_{-i}]^{-1} X_i^T &= g X_i [(1+g)X^TX - gX^T_{i} X_{i}]^{-1} X_i^T \\
&= g X_i \Big[\frac{(X^TX)^{-1}}{1+g} + \frac{g}{(1+g)^2} \frac{ (X^TX)^{-1} X_i^T X_i (X^TX)^{-1}}{1 - g(1+g)^{-1} h_{ii} } \Big] X_i^T \\
&= \frac{g}{1+g} \Big[ h_{ii} + \frac{gh_{ii}^2}{1 + g - g h_{ii}} \Big] \\
&= \frac{g}{1+g} \Big[ \frac{ h_{ii} + gh_{ii} - gh_{ii}^2 + gh_{ii}^2}{1 + g - g h_{ii}} \Big] \\
&= \frac{g h_{ii} }{1 + g(1 -  h_{ii} )}  .
\end{aligned}
$$
Therefore,
$$
\begin{aligned}
 E \big \{ \PVSI(Y_i \mid Y_{-i}; \theta) \mid Y_{-i} \big \} &= \frac{g^2}{(1+g)^2} h_{ii} E[\sigma^2 \vert Y_{-i} ](1 + g X_i [X^TX + gX^T_{-i} X_{-i}]^{-1} X_i^T) \\
 &= \frac{g^2}{(1+g)^2} h_{ii} E[\sigma^2 \vert Y_{-i} ]\Big(1 +  \frac{ g h_{ii} }{1 + g(1 -  h_{ii} )} \Big)  \\
 &= \frac{g^2}{(1+g)^2} h_{ii} E[\sigma^2 \vert Y_{-i} ]\Big( \frac{ 1 + g -  gh_{ii} + g h_{ii} }{1 + g(1 -  h_{ii} )} \Big) \\
 &= \frac{g^2}{(1+g)^2} h_{ii} E[\sigma^2 \vert Y_{-i} ] \frac{ 1 + g  }{1 + g(1 -  h_{ii} )} .
\end{aligned}
$$

Finally, we show the distribution of the EVOIR in linear regression with the g-prior. The predictive distribution of $Y_i$ is such that
$$
Y_i = X_i \mu_i + \sqrt{\frac{ \sigma_0^2 + Y_{-i}^T [ I_{n-1} + g X_{-i}(X^TX)^{-1}X_{-i}^T ]^{-1} Y_{-i}  }{ n (1 + X_i \Sigma_i X_i^T )^{-1} }} T_i
$$
where $\mu_i$ and $\Sigma_i$ are defined as in the previous section and $T_i \vert Y_{-i} \sim t_n$. So,
$$
\begin{aligned}
\hat{\beta}_{ls} &= (X^TX)^{-1} X^T Y \\
&= (X^TX)^{-1} X_{-i}^T Y_{-i} + (X^TX)^{-1} X_{i}^T Y_{i} \\
&= \hat{\beta}_{-i} + (X^TX)^{-1} X_{i}^T X_i \mu_i + \sqrt{\frac{ (n-2 ) \hat{\sigma}_{-i,g}^2 }{ n (1 + X_i \Sigma_i X_i^T )^{-1} }} (X^TX)^{-1} X_{i}^T  T_i.
\end{aligned}
$$
Since $E[T_i \vert Y_{-i}] = 0$,
$$
\hat{\beta}_{ls} = E[\hat{\beta}_{ls} \vert Y_{-i} ] + \sqrt{\frac{ (n-2 ) \hat{\sigma}_{-i,g}^2 }{ n (1 + X_i \Sigma_i X_i^T )^{-1} }} (X^TX)^{-1} X_{i}^T  T_i
$$

However, using a result from the previous section and the law of total expectation,
\begin{align*}
    E[\hat{\beta}_{ls} \vert Y_{-i} ] &= E[ \frac{1+g}{g} E[\beta \vert Y] \vert Y_{-i} ] \\
    &= \frac{1+g}{g}  E[ \beta \vert Y_{-i} ] \\
    &=  \Big(\frac{1+g}{g} \Big)  \Big( \frac{g}{g+1} \Big) \Big[ \hat{\beta}_{-i}-  \frac{1}{1+g + X_i (X^T_{-i} X_{-i} )^{-1} X^t_i}  (X^T_{-i} X_{-i} )^{-1}  X^T_iX_i \hat{\beta}_{-i} \Big] \\
    &= \hat{\beta}_{-i}-  \frac{1}{1+g + X_i (X^T_{-i} X_{-i} )^{-1} X^t_i}  (X^T_{-i} X_{-i} )^{-1}  X^T_iX_i \hat{\beta}_{-i} .
\end{align*}
It follows that,
$$
D_g(\hat{\beta}_{ls}, \hat{\beta}_{-i} ) = \sqrt{\frac{ (n-2 ) \hat{\sigma}_{-i,g}^2(1 + X_i \Sigma_i X_i^T )}{ n  }} (X^TX)^{-1} X_{i}^T  T_i
$$
Therefore,
\begin{align*}
    D_g(\hat{\beta}_{ls}, \hat{\beta}_{-i} ) ^T X^TX D_g(\hat{\beta}_{ls}, \hat{\beta}_{-i} ) &=  \frac{ (n-2 ) \hat{\sigma}_{-i,g}^2(1 + X_i \Sigma_i X_i^T )}{ n  } X_{i} (X^TX)^{-1} X_{i}^T  T_i^2 \\
    &= \frac{n-2}{n} \hat{\sigma}_{-i,g}^2(1 + X_i \Sigma_i X_i^T ) h_{ii} T_i^2 \\
\end{align*}
Note that, 
$$
Var(Y_i \vert Y_{-i}) =  \hat{\sigma}_{-i,g}^2 (1 + X_i \Sigma_i X_i^T ).
$$
So,
\begin{align*}
    D_g(\hat{\beta}_{ls}, \hat{\beta}_{-i} ) ^T X^TX D_g(\hat{\beta}_{ls}, \hat{\beta}_{-i} ) &=  \frac{ (n-2 ) \hat{\sigma}_{-i,g}^2(1 + X_i \Sigma_i X_i^T )}{ n  } X_{i} (X^TX)^{-1} X_{i}^T  T_i^2 \\
    &= Var(Y_i \vert Y_{-i}) \frac{n-2}{n}  h_{ii} T_i^2 \\
\end{align*}
So,
\begin{align*}
\EVOIR(Y_i \mid Y_{-i}) &= \frac{Var(Y_i \vert Y_{-i}) \frac{n-2}{n}  h_{ii} T_i^2 }{h_{ii} Var(Y_i \vert Y_{-i})} \\
&= \frac{n-2}{n} T_i^{2}
\end{align*}
Therefore,
$$
\EVOIR(Y_i \mid Y_{-i}) \vert Y_{-i}  \sim \frac{n-2}{n} F_{1,n}.
$$

\newpage
\section{Approximation of $E\{\pi \mid Y^{(k)}\}$ for Eswatini Example}
\begin{figure}[!htbp]	
  \begin{center}    
    \includegraphics[scale=.55,angle =270]{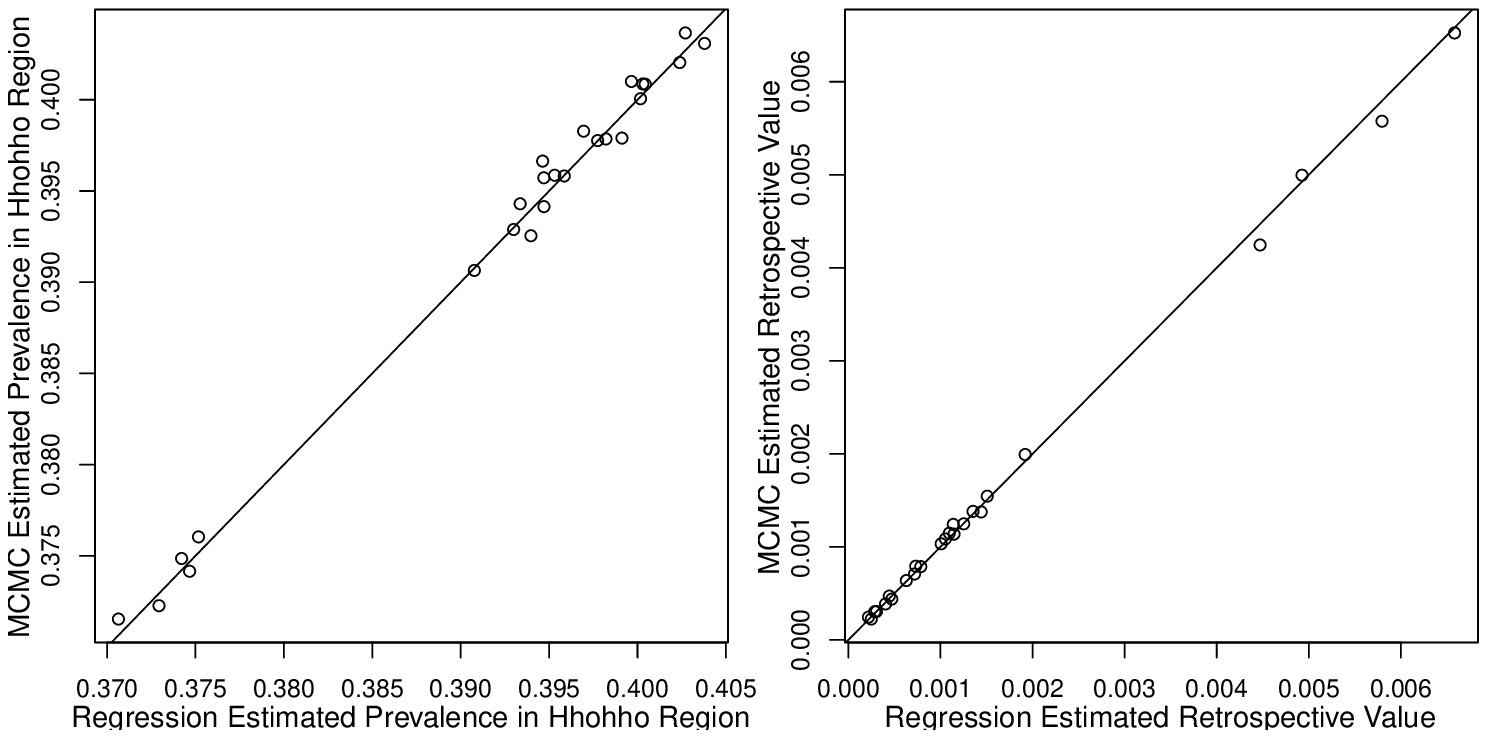} 
    \caption{The left plot shows the expected prevalence of HIV at each site estimated using an MCMC approach plotted against the expected prevalence as estimated by the $k$-nearest neighbors approach for multiple simulated complete data samples (filling in the removed FLAS clinic). The right plot shows the corresponding estimates for retrospective value of sample information for each method plotted against each other.}
    \label{fig:verification}      
  \end{center}
\end{figure}

We used a $k$ nearest neighbor regression to approximate $E\{\pi \mid Y^{(k)}\}$. In this approach, the expected prevalence conditional on a simulated complete data set is taken to be the average of the corresponding true prevalence of the 1000 nearest simulated complete data samples. Figure S\ref{fig:verification} indicates that there is a strong agreement between this approximation approach and the results that would occur by sampling from the posterior distribution conditional on the complete data directly to estimate the conditional means with high precision, having an MCMC standard error of less than $.000001$. The average error in the $\RVSI$ estimates is $.00002$, small enough that no conclusions based on this analysis would be changed if the results varied an amount on this order of magnitude.

\newpage
\section{Table of VOI Measures for Longley Data}
\begin{table}[!htbp]
  \centering
  \begin{tabular}{ccccc}
    Year & Cook's D     & $\RVSI$ & $\PVSI$ &  $\EVOIR$ \\
    \hline
    1947 & 0.141 & 0.092                & 0.088 &1.05        \\
    1948 & 0.041 & 0.026                &  0.177 &0.15     \\
    1949 & 0.003 & 0.002                & 0.079& 0.02     \\
    1950 & 0.244 & 0.159                & 0.056& 2.83     \\
    1951 & 0.614 & 0.399                & 0.157& 2.55     \\
    1952 & 0.089 & 0.058                &  0.072& 0.80     \\
    1953 & 0.079 & 0.051                & 0.126& 0.41       \\
    1954 & 0.001 & 0.000                & 0.142 & 0.00    \\
    1955 & 0.000 & 0.000                &  0.117& 0.00   \\
    1956 & 0.235 & 0.153                & 0.043& 3.53   \\
    1957 & 0.000 & 0.000                & 0.078& 0.00      \\
    1958 & 0.004 & 0.002                & 0.130& 0.02      \\
    1959 & 0.036 & 0.023                & 0.080& 0.29       \\
    1960 & 0.004 & 0.003                &  0.041& 0.07      \\
    1961 & 0.170 & 0.111                & 0.064&1.72     \\
    1962 & 0.467 & 0.304                & 0.258 & 1.18    
  \end{tabular}
  \label{tab:longleyVOI}
  \caption{VOI Measures for Longley Data: Cook's distance, the retrospective expected value of sample information ($\RVSI$), the prospective expected value of sample information ($\PVSI$), and the expected value of information ratio ($\EVOIR$).}
\end{table}

\newpage
\section{HIV Prevalence Estimates and VOI Measures for Eswatini Example}

\begin{figure}[t!]  	
  \begin{center}    
    \includegraphics[scale=.65, angle = 270]{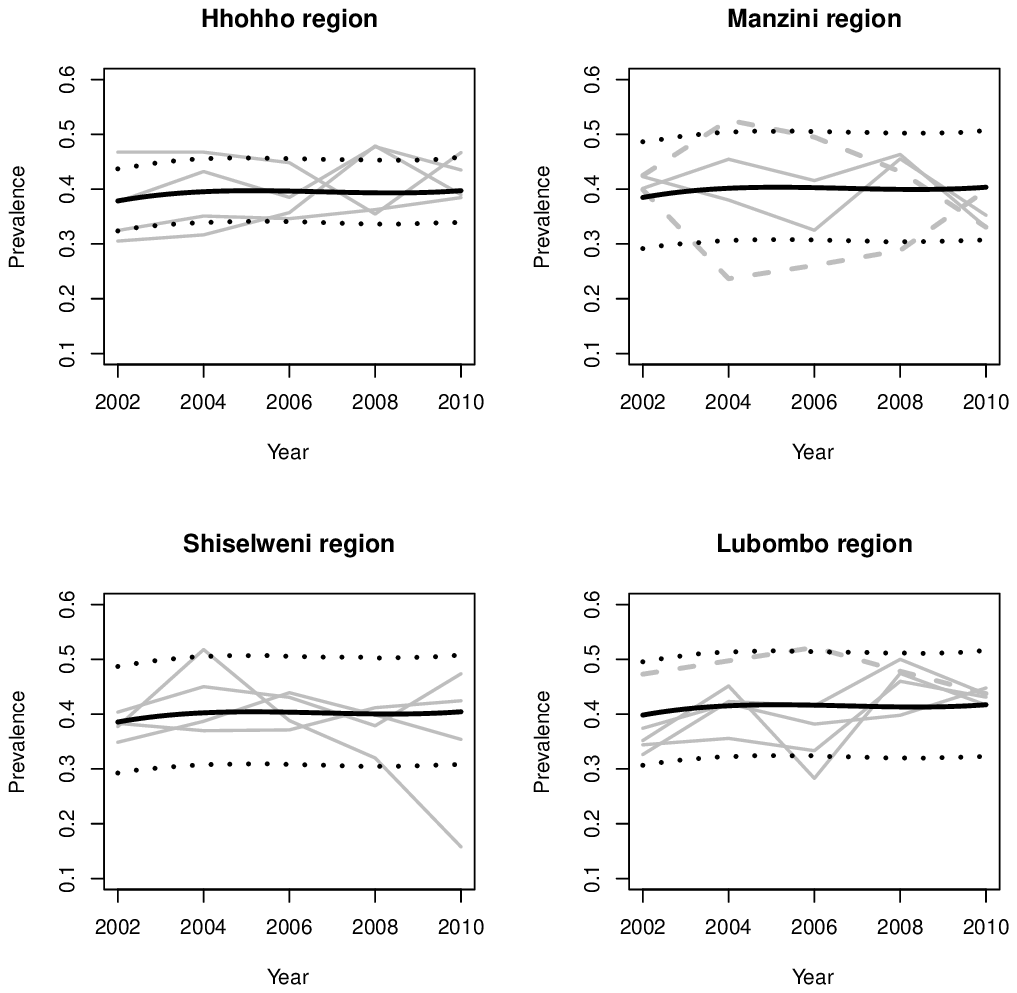} 
    \caption{The fitted HIV prevalence curve for each region is shown as a bold black line with the bold dashed lines indicating pointwise $95\%$ credible intervals. The gray lines are the observed prevalence at each site. The FLAS Clinic, Vuvulane Clinic, and the King Sobhuza II PHU have prevalence indicated by dashed gray lines.}
    \label{fig:fitted}      
  \end{center}     
\end{figure}

Figure S\ref{fig:fitted} presents the fitted HIV prevalence curve for each region as a bold black line with the bold dashed lines indicating pointwise $95\%$ credible intervals. Also shown in Figure \ref{fig:fitted}, are the observed percentages of individuals that tested positive for HIV at each site. The HIV prevalence rates are fairly stable in 2000's as being observed in most clinics. The posterior medians are at similar level across four regions. Hhohho region estimates have a smaller uncertainty than other three regions because clinics in this region have similar trends. 

Table \ref{tab:SwazilandVOI} presents the three value of information measures. Both the prospective and the retrospective expected value of sample information have a Monte Carlo standard error of less than .0001 for each site.

\begin{table}[!htbp]
  \centering
  \begin{tabular}{llccc}
    Clinic  & Region     & $\PVSI$ & $\RVSI$ &  $\EVOIR$ \\
    \hline
    Mbabane       &   Hhohho    & 0.0015 & 0.0007 & 0.44 \\
    Piggs Peak      &   Hhohho    & 0.0012 & 0.0012 & 1.03  \\
    Mkhuzweni HC     &   Hhohho    & 0.0015 & 0.0014 & 0.92 \\
    Dvokolwako      &   Hhohho    & 0.0014 & 0.0008 & 0.57 \\
    King Sobhuza II PHU &  Manzini    & 0.0013 & 0.0023 & 1.69 \\
    FLAS Clinic     &  Manzini    & 0.0012 & 0.0029 & 2.40 \\
    Mankayane HC     &  Manzini    & 0.0015 & 0.0002 & 0.16 \\
    Luyengo Clinic    &  Manzini    & 0.0014 & 0.0007 & 0.45 \\
    Hlathikhulu PHU   & Shiselweni  & 0.0015 & 0.0002 & 0.14 \\
    Nhlangano HC     & Shiselweni  & 0.0016 & 0.0006 & 0.38 \\
    Matsanjeni HC    & Shiselweni  & 0.0015 & 0.0001 & 0.06 \\
    Dwaleni Clinic    & Shiselweni  & 0.0011 & 0.0001 & 0.07 \\
    Siteki PHU      &  Lubombo    & 0.0009  & 0.0001 & 0.11 \\
    Lomahasha Clinic   &  Lubombo    & 0.0010 & 0.0001 & 0.13 \\
    Sithobela HC     &  Lubombo    & 0.0008 & 0.0008  & 0.96  \\
    Ndevane Clinic    &  Lubombo    & 0.0010  & 0.0001 & 0.10 \\
    Vuvulane Clinic   &  Lubombo    & 0.0007 & 0.0016 & 2.31
  \end{tabular}
  \label{tab:SwazilandVOI}
  \caption{VOI Measures for Eswatini Example: the retrospective expected value of sample information ($\RVSI$), the prospective expected value of sample information ($\PVSI$), and the expected value of information ratio ($\EVOIR$).}
\end{table}

\vspace{-0.3in}
\bibliographystyle{chicago}      
\bibliography{citations}   